\documentclass[final,1p,times]{elsarticle}

\usepackage{graphicx}
\usepackage{amssymb}
\usepackage{ulem}
\usepackage[usenames]{color}
\usepackage{amsmath,amssymb,mathrsfs}
\usepackage{dsfont}

\newcommand{\be}{\begin{eqnarray}}
\newcommand{\ee}{\end{eqnarray}}
\newcommand{\non}{\nonumber \\}
\newcommand{\po}{{\rm P}}
\newcommand{\npo}{{\rm NP}}

\newcommand{\oh}{\frac{1}{2}}
\newcommand{\Nst}{N^{*}}
\newcommand{\Dst}{\Delta^{*}}
\newcommand{\mpi}{m_{\pi}}
\newcommand{\gam}{\gamma^{\mu}}
\newcommand{\gan}{\gamma^{\nu}}
\newcommand{\vtau}{\vec{\tau}}
\newcommand{\delpi}{\partial_{\mu}\vec{\pi}}
\newcommand{\gaf}{\gamma^{5}}
\newcommand{\delmu}{\partial_{\mu}}
\newcommand{\delnu}{\partial_{\nu}}
\newcommand{\kst}{K^{*}}
\newcommand{\Sigst}{\Sigma^{*}}

\journal{Nuclear Physics A}

\begin{document}
\begin{frontmatter}

\author[juli]{M.~D\"oring}
\ead{m.doering@fz-juelich.de}

\author[juli]{C.~Hanhart}

\author[uga]{F.~Huang}

\author[juli]{S.~Krewald} 

\author[bonn,juli]{U.-G.~Mei\ss ner}

\author[juli]{D.~R\"onchen}

\address[juli]{Institut f\"ur Kernphysik and J\"ulich Center for Hadron Physics, 
Forschungszentrum J\"ulich, D-52425 J\"ulich,Germany}
\address[uga]{Department of Physics and Astronomy, University of Georgia, Athens, Georgia 30602, USA}
\address[bonn]{Helmholtz-Institut f\"ur Strahlen- und Kernphysik (Theorie) and Bethe Center for Theoretical
Physics, \\
Universit\"at Bonn, Nu\ss allee 14-16, D-53115 Bonn, Germany
}
\title{
 \hspace{8.1cm}{\footnotesize FZJ-IKP-TH-2010-21, HISKP-TH-10/25}
\\
The reaction $\pi^+p\to K^+\Sigma^+$ in a unitary coupled-channels model
}

\begin{abstract}
Elastic $\pi N$ scattering and the reaction $\pi^+ p \to K^+\Sigma^+$ are 
described simultaneously in a
unitary coupled-channels approach which respects analyticity. SU(3) flavor symmetry is used to relate the
$t$- and $u$- channel exchanges that drive the meson-baryon interaction in the different channels. Angular
distributions, polarizations, and  spin-rotation parameters are compared with available experimental data.
The pole structure of the amplitudes is extracted from the analytic continuation. 
\end{abstract}


\end{frontmatter}


\section{Introduction}

The excitation spectrum of baryons and mesons is expected to reveal important information on the mechanism
of confinement as well as the intrinsic structure of hadrons. Properties of baryon resonances have been
obtained by lattice calculations ~\cite{Durr:2008zz,Dudek:2009qf,Bulava:2010yg,Engel:2010my}, mostly for the
ground states but also for some excited states~\cite{Dudek:2009qf,Bulava:2010yg}.  In quark
models~\cite{Isgur:1978xj,Capstick:1986bm,Loring:2001kx},  a rich spectrum of excited states is predicted.
Many of these resonances    could be identified in elastic $\pi N$ scattering,  while at higher energies,
usually more states are predicted than seen, a fact commonly referred to as the ``missing resonance
problem''~\cite{Koniuk:1979vw}.  Since resonances not seen in the $\pi N$ channel might predominantly couple
to other channels, there are intensive experimental efforts~\cite{Klempt:2009pi} to measure, among others,
multi-pion or $KY$ final states, where $KY=K\Lambda$ or $K\Sigma$.

 The reaction $ \pi^+  p \rightarrow K^+ \Sigma^+$ provides access to a pure isospin $I=3/2$ two-body
reaction channel in meson-nucleon dynamics. Moreover, the weak decay $\Sigma^+ \rightarrow p \pi^0$ 
allows to determine the polarization of the produced $ \Sigma^+ $. In the
1980's, Candlin {\it et al.}  measured differential cross sections and polarizations at the Rutherford
Appleton Laboratory for pion beam momenta ranging from $p_{\pi} = 1.282 $ GeV/c to $p_{\pi} = 2.473 $
GeV/c~\cite{Candlin:1982yv} and  performed an energy-dependent isobar analysis for invariant collision
energies ranging from the $ K^+ \Sigma^+$ threshold ($z\equiv\sqrt{s}=1.68$ GeV) to $\sqrt{s}=2.35$
GeV~\cite{Candlin:1983cw}. While the quality of the fit is good, unitarity is violated and a separation of
the resonant part is difficult due to the oversimplified construction of the partial wave amplitudes. The
resonance parameters extracted confirmed four $\Delta$-resonances found previously in the partial wave
analyses of elastic pion-nucleon scattering by Cutkosky {\it et al.}~\cite{Cutkosky} and    H\"ohler {\it et
al.}~\cite{Hoehler1,Hoehler2},  the $\Delta(1905)F_{35}$,  $\Delta(1920)P_{33}$, $\Delta(1950)F_{37}$, and
$\Delta(2200)G_{37}$. Other resonances deduced from elastic pion-nucleon scattering by
Refs.~\cite{Cutkosky,Hoehler1,Hoehler2}  could not be unambiguously identified in
$K^+\Sigma^+$ production~\cite{Candlin:1983cw}, notably the  $\Delta(1910)P_{31}$, which is given a
4-star status by the PDG, and the  $\Delta(1900)S_{31}$, downsized to a one star rating
(nowadays, two star) after the $K\Sigma$ data became available.

In 1988, Candlin {\it et al.}~\cite{Candlin:1988pn} measured the spin-rotation parameter $\beta$ at
the CERN-SPS, using a polarized frozen spin target and the Rutherford Multiparticle Spectrometer RMS, adding
independent information to the data base for two pion beam momenta  $p_{\pi} = 1.69 $~GeV/c and $p_{\pi} =
1.88 $~GeV/c. The spin-rotation parameter correlates the spin of the target proton and the spin of
the produced $\Sigma^+$ and allows to eliminate ambiguities in the partial wave analysis. Discrepancies
between the $\beta$-values predicted from the partial wave analysis of Ref.~\cite{Candlin:1983cw} and the
experimental ones were found which suggested the necessity of a new partial wave
analysis~\cite{Candlin:1988pn}.

A consolidated knowledge of coupled-channels meson-baryon (MB) scattering is required when studying meson
production in $NN$ collisions, such as measured at COSY/J\"ulich~\cite{Valdau:2010kw,AbdElSamad:2010tz}.
There, the $MB\to MB$ transitions enter the proton induced strangeness production~\cite{Valdau:2010kw,
Gasparyan:2003cc} as sub-processes in on-shell but also off-shell kinematics.

Also, a detailed knowledge of the resonance content in $K\Sigma$ production is needed in heavy  ion
collisions. The $K^+$ has a long mean free path in the nucleus and is believed to provide information  about
the high density and temperature phase of the heavy ion  collision~\cite{Tsushima:1994rj}; to clarify the
role of the $\Delta(1920)P_{33}$ in the $\pi^+p\to K^+\Sigma^+$ reaction is thus
mandatory~\cite{Tsushima:1994rj}.

Various analyses of meson-baryon scattering are available, designed with the goal to extract resonance
properties from data. They differ, e.g., in the number of channels and their analytic properties. Some
representative analyses are discussed in the following.

A coupled reaction channel analysis of nucleon resonances including the $K\Sigma$ channel has been performed
by the Gie{\ss}en group in the $K$-matrix approximation~\cite{Penner:2002ma,Penner:2002md}. Elastic $\pi N$
scattering, as well as the family of $\pi N\to KY,\,\eta N,\,\omega N$ and other reactions are included in
the analysis. Photon-induced reactions within the model have been studied in Ref.~\cite{Penner:2002md}.  The
non-resonant part of the amplitude is treated in a Lagrangian approach and resonances are included up to a
total spin of $J=3/2$. In more recent studies~\cite{Shklyar:2004dy,Shklyar:2009cx}, the spin $5/2$
resonances have been included within a Lagrangian-based framework. 
Unitarity is respected, but the real, dispersive parts of the
two-body intermediate states are neglected, such that analyticity is lost.

While in this analysis the imaginary part from phase space is cut off at threshold,  in other approaches it
is analytically continued below threshold, but the dispersive parts are still not included. Such $K$-matrix
approaches~\cite{Manley:1992yb,Scholten:1996mw} analyze $\pi N$, $\eta N$ or $\pi\pi N$ data, or even more
reactions like the Bonn-Gatchina  group~\cite{Anisovich:2004zz,Sarantsev:2005tg,Anisovich:2010an}.  

A very precise analysis of elastic $\pi N$ scattering is provided by
the $K$-matrix approach of the GWU/SAID group~\cite{Arndt:1995bj,Arndt:2003if,Arndt:2006bf,Workman:2008iv}.
There are no assumptions made about resonances [except for the $\Delta(1232)$] and in this sense the
extraction of the excited baryon spectrum is model-independent.  Also, this partial wave analysis provides
the lowest $\chi^2$ of the available analyses of elastic $\pi N$ scattering~\cite{Arndt:2006bf}. This is
also the reason, why in this study  we use the analysis of Ref.~\cite{Arndt:2006bf} as input rather than
directly fitting to $\pi N$ data, although a direct fit to data should be carried out in the future.
Interestingly, in the most recent update of the analysis~\cite{Arndt:2006bf}, several resonances with less
than four stars could not be confirmed any more.

Carnegy-Mellon-Berkeley (CMB) type of models~\cite{Batinic:1995kr,Vrana:1999nt,Ceci:2008zz} usually include
the dispersive parts of the resonance propagators but do not provide a microscopical background.

Dynamical coupled-channels models take the real, dispersive parts of the intermediate states into account
and provide a microscopical description of the background~\cite{Surya:1995ur, Meissner:1999vr,
Schutz:1998jx, Krehl:1999km, Gasparyan:2003fp, Doring:2009yv, Matsuyama:2006rp, Chen:2007cy, Suzuki:2008rp,
Wagenaar:2009gr}.
Dynamical coupled-channels models are based on effective Lagrangians. While $\pi N$ scattering at low
energies is completely understood from chiral perturbation theory (see, e.g., Refs.~\cite{Fettes:1998ud,
Fettes:2001cr} or  Refs.~\cite{Meissner:1999vr, Gasparyan:2010xz} for unitarized extensions of $\chi$PT), at
higher energies model assumptions need to be made. It is realistic to assume that the interaction is driven
by the exchange of known mesons and baryons. The  scattering amplitudes are then obtained as solutions of a
Lippmann-Schwinger equation which guarantees unitarity. Thus, the driving term of the Lippmann Schwinger
equation consists of $t$-channel meson exchange processes and $u$-channel baryon exchanges as well as
$s$-channel processes which may be considered as bare resonances. 

The explicit treatment of the $t$-channel and $u$-channel diagrams introduces strong correlations between
the different partial waves and may generate a non-trivial energy and angular dependence of the observables.
The explicit treatment of this background in terms of exchange diagrams also allows to link different
reactions such as elastic $\pi N$ scattering and the reaction $\pi^+ p\to K^+\Sigma^+$, using SU(3) flavor
symmetry. Thus, the treatment of the interaction via meson and baryon exchange is expected to lead to a
realistic background, with strong restrictions on the free parameters.

In view of this, the strategy to perform baryon spectroscopy is to introduce only a minimum number of bare
resonance states in order to obtain a good description of the data. This distinguishes the ansatz from some
$K$-matrix approaches where the absence of a structured background may require the introduction of
additional resonance states, which improve the $\chi^2$ but are in fact simulating the background.

Dynamical coupled-channels approaches have been so far restricted to the reaction channels $N \pi, N\eta,
N\sigma, \Delta \pi$, and $N \rho$ \cite{Krehl:1999km, Gasparyan:2003fp, Suzuki:2008rp}, and concentrated on
differential cross sections, mostly of elastic $\pi N$ scattering. In the present study, we extend the
dynamical coupled-channels {\it J\"ulich model}, which has been developed over the
years~\cite{Schutz:1998jx, Krehl:1999km, Gasparyan:2003fp, Doring:2009yv}, to the kaon-hyperon sector by
adding Lagrangians for the couplings to the kaon hyperon channels and  resonances beyond the set considered
in Refs.~\cite{Schutz:1998jx, Krehl:1999km, Gasparyan:2003fp, Doring:2009yv}. We limit our resonance
analysis to the energy range investigated in Refs.~\cite{Schutz:1998jx, Krehl:1999km, Gasparyan:2003fp,
Doring:2009yv}, i.e. 2 GeV, and concentrate on the isospin $I=3/2$ sector.

In Sec.~\ref{sec:newpotentials} the inclusion of the $KY$ channels in addition to the channels $N \pi,
N\eta, N\sigma, \Delta \pi$, and $N \rho$ is discussed. To describe the data in the $\pi^+p\to K^+\Sigma^+$
reaction, we also need to include higher spin resonances up to a total spin of $J=7/2$ (cf.
Sec.~\ref{sec:schannel}). Results are presented in Sec.~\ref{sec:1}. For the analysis of the resonance
content of the resulting amplitude, given by the pole positions and residues,  one needs the analytic
continuation, summarized in Sec.~\ref{sec:analytisch}. The extracted resonance properties are listed and
commented on in Secs.~\ref{sec:poles} and \ref{sec:branching}. In \ref{sec:su3} (\ref{sec:bare_res}), 
the $t$- and $u$- ($s$-)channel processes used in this study are explicitly given.


\section{Formalism}
\subsection{Scattering equation}
The coupled-channels scattering
equation~\cite{Schutz:1998jx,Krehl:1999km,Gasparyan:2003fp,MuellerGroeling:1990cw} used in the present
formalism fulfills  two-body unitarity, as well as some requirements of three-body unitarity following
Ref.~\cite{Aaron:1969my}.  Furthermore, it fulfills analyticity and takes into account the dispersive parts
of the intermediate states as well as the off-shell behavior dictated by the interaction  Lagrangians. This
integral equation which is solved in the $JLS$-basis is given by
\be
&&\langle L'S'k'|T_{\mu\nu}^{IJ}|LSk\rangle=\langle L'S'k'|V_{\mu\nu}^{IJ}|LSk\rangle\non
&&+\sum_{\gamma\, L''\, S''}\int\limits_0^\infty k''^2\,dk''
\langle L'S'k'|V_{\mu\gamma}^{IJ}|L''S''k''\rangle
\,\frac{1}{z-E_{\gamma}(k'')+i\epsilon}\,\langle L''S''k''|T_{\gamma\nu}^{IJ}|LSk\rangle
\label{bse}
\ee
where $J\,(L)$ is the total angular (orbital angular) momentum, $S\,(I)$ is the total spin (isospin),
$k(k',\,k'')$ are the incoming (outgoing, intermediate) momenta,  and $\mu,\,\nu,\,\gamma$ are channel
indices. The incoming and outgoing momenta can be on- or off-shell. In Eq.~(\ref{bse}),
$E_\gamma$ is the on-mass shell energy in channel $\gamma$, 
$E=\sqrt{m^2+(k'')^2}+\sqrt{m_B^2+(k'')^2}$
where $m\, (m_B)$ is the meson (baryon) mass. The second term in
Eq.~(\ref{bse}) on the right-hand side involves also a sum over all intermediate possible quantum numbers
and channels contained in the model.

For the channels involving quasi-particles, $\sigma N$, $\rho N$, and $\pi\Delta$, the propagator is
slightly more complicated~\cite{Schutz:1998jx, Krehl:1999km, Krehl:1999ak} (cf. also 
Sec.~\ref{sec:analytisch}). The pseudo-potential $V$ iterated in Eq. (\ref{bse}) is constructed from an
effective interaction based on the Lagrangians of Wess and Zumino \cite{Wess:1967jq,Meissner:1987ge},
supplemented by additional terms \cite{Krehl:1999km,Gasparyan:2003fp} for including the $\Delta$ isobar, the
$\omega$, $\eta$, $a_0$ meson, and the $\sigma$ [cf. Sec.~\ref{sec:newpotentials}]. The exchange potentials
$V$ are partial wave projected to the $JLS$-basis.

The novelty in this work is the inclusion of the $KY$ channels $K\Lambda$ and $K\Sigma$. This leads to a
larger channel space and new transition potentials $V$ to and within the $KY$ channels. These new potentials
$V$, related to the existing ones by SU(3) symmetry, are discussed in Sec.~\ref{sec:newpotentials}.
They also contain the form factors which are used to regularize the scattering equation (\ref{bse}).


\subsection{$s$-channel processes}
\label{sec:schannel}
In a model with explicit $s$-channel states it is always possible to separate the amplitude into a pole and
a non-pole part 
\be
T=T^\po+T^\npo
\label{deco1}
\ee 
where the pole part $T^\po$ is defined as the set of diagrams that is 1-particle reducible, i.e. there is at
least one $s$-channel exchange. Usually, the non-pole, 1-particle irreducible part $T^\npo$ comes from  $t$-
and $u-$ channel exchange processes collected into the non-pole potential $V^{\npo}$ which is then
unitarized using a dynamical equation  of the type of Eq.~(\ref{bse}) --- see also Eq.~(\ref{dressed})
below. The separation of the type of Eq.~(\ref{deco1}) is widely used in the literature, see e.g.
\cite{Matsuyama:2006rp,Afnan:1980hp}. $T^\npo$ is usually referred to as {\it background}, although the
unitarization may lead to dynamically generated poles in $T^\npo$ as discussed in detail in
Ref.~\cite{Doring:2009bi}. There, the conclusion was drawn that the clearest separation into a background
and a resonance part is given by the separation into a singularity-free part and the part $a_{-1}/(z-z_0)$
from the leading term in the Laurent expansion [cf. Eq.~(\ref{pa})].

In the present study, we use the decomposition of Eq.~(\ref{deco1}), because the calculation of $T^\po$ is
numerically much faster than that of $T^\npo$. In a fit of only $s$-channel parameters, it is thus
convenient to calculate $T^\npo$ once and then fit the resonance parameters, which only requires the
multiple re-evaluation of $T^\po$. Note that resonance $u$-channel exchanges contribute  to all partial
waves and are thus accounted to $T^\npo$. Nucleon, $\Lambda$, $\Sigma$, $\Delta(1232)$, and 
$\Sigma^*(1385)$ $u$-channel exchange diagrams are included with physically known coupling strengths [see
\ref{sec:su3}],  while $u$-channel diagrams from other baryonic resonances are neglected. Those would
introduce additional parameters which are difficult to adjust for the diagrams do not introduce strong
energy dependencies (for a discussion of $u$-channel contributions see Ref.~\cite{Lutz:2001yb}).

The pole contribution $T^\po$ can be evaluated from the non-pole part $T^\npo$, i.e. from the set of
diagrams that is 1-particle irreducible. For this, we define the following quantities in a given partial
wave, 
\be
T^\npo(d,c)&=& V^{\npo}(d,c)+V^{\npo}(d,e)G(e)T^\npo(e,c) \non
\Gamma_D^{(\dagger)}(i,c)&=&\gamma_B^{(\dagger)}(i,c)+\gamma_B^{(\dagger)}(i,d)\,G(d)\,T^\npo(d,c)\non
\Gamma_D(c,i)&=&\gamma_B(c,i)+T^\npo(c,d)\,G(d)\,\gamma_B(d,i)\non
\Sigma(i,j)&=&\gamma_B^{(\dagger)}(i,c)\,G(c)\,\Gamma_D(c,j)
\label{dressed}
\ee 
where $\Gamma_D^{(\dagger)}$ ($\Gamma_D$) are the dressed resonance creation (annihilation) vertices and
$\Sigma$ is the self-energy. The indices $i,j$ indicate the resonance in the case of multiple resonances,
while $c,\,d,\,e$ are indices in channel space. Integrals and sums over intermediate states are not
explicitly shown in Eq. (\ref{dressed}). 

For the two-resonance case, the pole part reads explicitly~\cite{Doring:2009uc}
\be
T^{\po}=\Gamma\, D^{-1} \, \Gamma^{(\dagger)},\quad
\Gamma=(\Gamma_1,\Gamma_2), 
\quad \Gamma^{(\dagger)}=\left(
\begin{matrix}
\Gamma^{(\dagger)}_1\\
\Gamma^{(\dagger)}_2
\end{matrix}
\right),\quad
D=\left(
\begin{matrix}
z-m_1-\Sigma_{11}&&-\Sigma_{12}\\
-\Sigma_{21}     &&z-m_2-\Sigma_{22}
\end{matrix}
\right)
\label{2res}
\ee
from which the one-resonance case follows immediately.  The bare vertices $\gamma_B$ for resonances with
Spin $J\le 3/2$ are derived from Lagrangians. The vertex functions for $J\geq 5/2$ are given in
Eq.~(\ref{higher1}). Further details on the $s$-channel processes are given in \ref{sec:bare_res}. 


\subsection{$t-$ and $u$-channel exchange processes}
\label{sec:newpotentials}
\begin{figure}
\begin{center}
\includegraphics[height=0.26\textwidth]{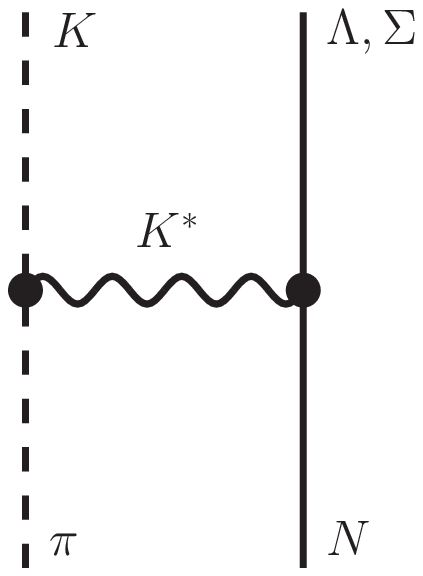} 
\includegraphics[height=0.26\textwidth]{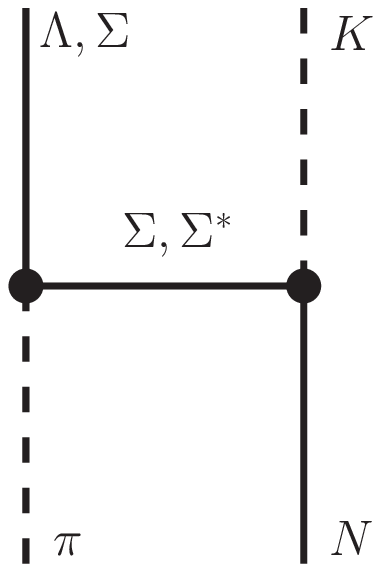}  
\includegraphics[height=0.26\textwidth]{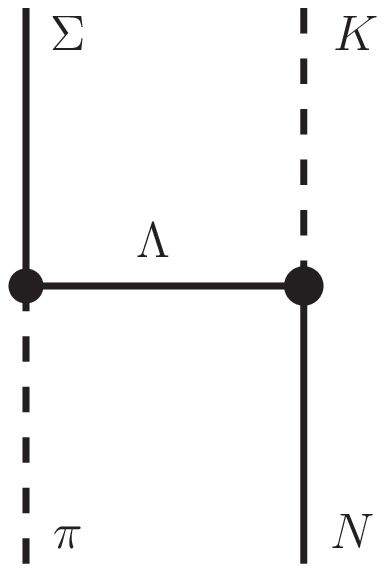} \\
\includegraphics[height=0.26\textwidth]{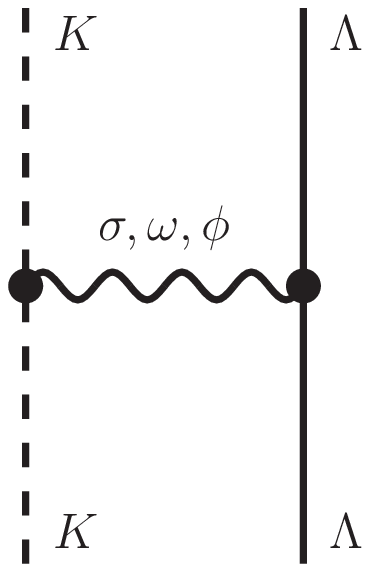}   
\includegraphics[height=0.26\textwidth]{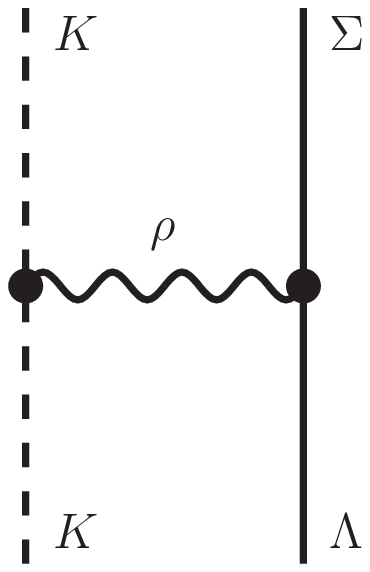}  
\includegraphics[height=0.26\textwidth]{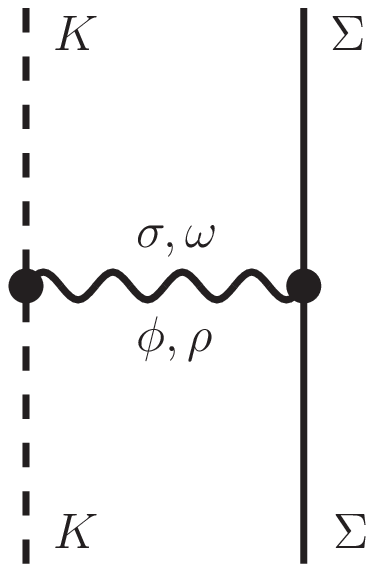}
\end{center}
\caption{$\pi N \to KY$ and $KY\to KY$ transitions. For the other transitions used in this study, 
see Refs.~\cite{Krehl:1999km,Gasparyan:2003fp}.}
\label{fig:kydiagrams}     
\end{figure}

The $t$- and $u$-channel processes provide the non-resonant interaction in the meson exchange picture. 
The transition potentials without participation of $KY$ have been derived in
Refs.~\cite{Schutz:1998jx,Krehl:1999km,Gasparyan:2003fp} and explicit expressions can be found in these
references. Here, we quote only the extension to the $KY$ channels. The corresponding exchange processes are
shown in Fig.~\ref{fig:kydiagrams}.

The vertices present in these diagrams are related to the already existing ones without strange particles
using SU(3) symmetry (except for the $\sigma$ meson, cf. \ref{sec:su3}). The coupling of SU(3) octets
depends on two parameters which can be related to the axial coupling and an additional parameter. The values
for these parameters have been taken from the literature and are not fitted in this study. This is explained
in detail in \ref{sec:su3}. There, one can also find the explicit amplitudes for the diagrams shown in
Fig.~\ref{fig:kydiagrams}. 

SU(3) symmetry is broken in the present study by the use of physical meson and baryon masses, as well as by
different cut-offs in the form factors of the vertices. Exchange processes with
strangeness $S=-2$ particles have been neglected because these baryons and, moreover, the corresponding
3-particle intermediate states, are heavy. 
A $\kappa$ exchange is in principle possible 
but not required by the data and thus has been neglected for simplicity.
Furthermore, $\rho N, \,\pi\Delta \leftrightarrow KY$ $t$- and $u$-channel transitions are neglected in the
present work, as they appear only at loop order in the considered reactions $\pi N\to\pi N$
and $\pi^+p \to K^+\Sigma^+$.


\subsection{Analytic continuation}
\label{sec:analytisch}
As argued in Ref.~\cite{Doring:2009bi}, a clean separation of resonances and background is possible by the
extraction of pole contributions from the analytic continuation. First results within different dynamical
coupled channels models have been obtained in Refs.~\cite{Doring:2009yv, Suzuki:2008rp, Tiator:2010rp}.

\begin{figure}
\begin{center}
\includegraphics[width=0.94\textwidth]{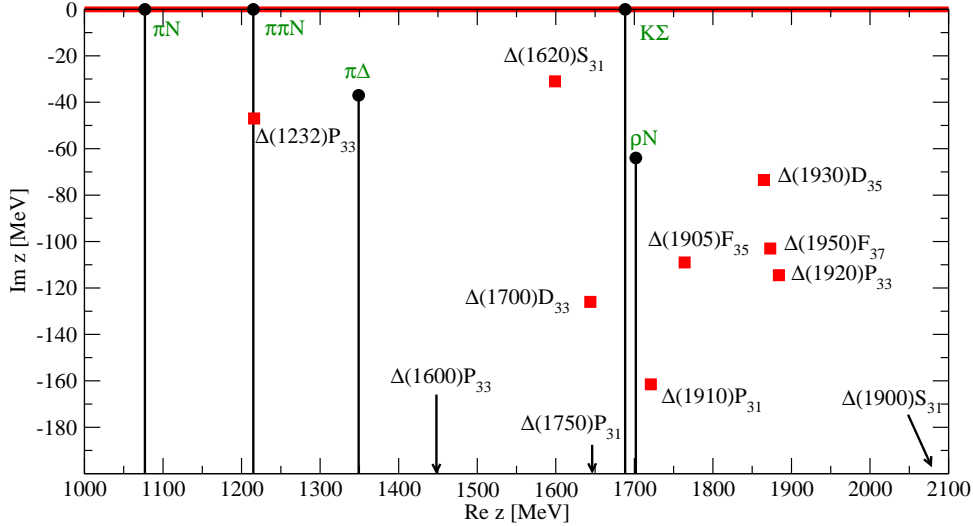} 
\end{center}
\caption{Branch points of the coupled channels and the chosen directions of the associated cuts. 
Also, the isospin $I=3/2$ pole positions on the second sheet $T^{(2)}$ are shown 
[cf. Sec.~\ref{sec:poles}].}
\label{fig:plane}     
\end{figure}

The analytic continuation of the amplitude within the present framework has been derived in
Ref.~\cite{Doring:2009yv} in detail. Here, we summarize only the analytic structure. For the channels with
stable two-body intermediate states, $\pi N$, $\eta N$, $K\Lambda$, and $K\Sigma$, there is one
branch point at threshold $z_{\rm thres}=m+m_B$ which induces one new sheet. This is called the unphysical
sheet. To search for poles on this sheet, it is convenient to rotate the right-hand, physical cut that
extends from $z_{\rm thres}$ to $\infty$, into the negative $\rm{Im}\, z$ direction as shown in
Fig.~\ref{fig:plane}. 
Poles on this redefined sheet are close to the physical axis. Poles on other sheets are situated further
away from the physical axis and thus typically have a much reduced effect on observables. However, there are
certain situations in which such ``shadow poles'' can cause structures on the physical axis. An example is
the $N^*(1535)$. There, the interplay between the usual pole and the shadow pole causes the $\eta N$ cusp
seen in the S11 partial wave, at least within the model of Ref.~\cite{Doring:2009yv}.

For the effective $\pi\pi N$ propagators $\pi\Delta,\,\sigma N$, and $\rho N$, the analytic structure is
more complicated: there is a branch point at $z_{\rm thres}=2m_\pi+m_N$ which is induced by the cut of the
self-energy of the unstable particle. Additionally, there are branch points in the complex plane at
$z'_{\rm thres}$ and $(z'_{\rm thres})^*$ with $z'_{\rm thres}=z_0+M$ where $M$ is the mass of the
stable particle and $z_0$ is the pole position in the scattering problem of the unstable particle in the
rest frame of the unstable particle~\cite{Doring:2009yv}. Those  branch points can be regarded as
pseudo-thresholds that have moved into the complex plane due to the unstable character of one of the
particles. The argument to chose the direction of the cuts, associated with the branch points of the
effective $\pi\pi N$ channels, is  the same as before: the cut is rotated into the negative ${\rm Im}\, z$
direction so that only those poles are found which are physically relevant. This is also indicated in
Fig.~\ref{fig:plane}.

For the effective $\pi\pi N$ channels, there are again situations, where poles on hidden sheets may have an
effect on the physical axis: In Ref.~\cite{Doring:2009yv} a state in $T^\npo$ in the $D_{13}$ partial wave
has been found, dynamically generated from the $S$-wave $\rho N$ interaction (cf. also
Refs.~\cite{Oset:2009vf,Lutz:2001mi}). This state, while its pole is well below the $\rho N$ branch point,
is visible as a washed-out structure at the nominal $\rho N$ threshold, due to the fact that there is no
direct connection from the pole to the physical axis, but only around the branch point at $z'_{\rm thres}$
in the complex plane. This resonance-like structure around $z=1700$ MeV in the $D_{13}$ partial wave is,
however, only visible in $T^\npo$; once the strong $N^*(1520)D_{13}$ resonance is included in the same
partial wave, resonance repulsion~\cite{Doring:2009bi} pushes the pole far into the complex
plane~\cite{Doring:2009yv}, so that we cannot identify the dynamically generated structure with the three
star $N^*(1700)D_{13}$ resonance~\cite{pdg}.

Another example is the Roper resonance $N^*(1440)$ whose poles are close to the $\pi\Delta$ pseudo-threshold
in the complex plane. In case the coupling of the Roper to the $\pi\Delta$ channel is large, this interplay
of usual and hidden pole with the branch point may lead to the non-trivial structure of the Roper resonance
visible on the physical axis~\cite{Arndt:2006bf,Doring:2009yv,Suzuki:2009nj}.

The second sheet of the amplitude $T$ denoted by $T^{(2)}$ in the following is shown in
Fig.~\ref{fig:plane}, with the cuts as defined above.  In order to extract a pole residue on $T^{(2)\,i\to
f}$ for a transition from channel $i$ to $f$, we expand the amplitude $T^{(2)}$ in a Laurent series around
the pole position, 
\be
T^{(2)\,i\to f}&=&\frac{a_{-1}^{i\to f}}{z-z_0}+a_0^{i\to f}+{\cal O}(z-z_0).
\label{pa}
\ee
In \ref{sec:residues}, the calculation of residues and branching ratios is discussed in detail, cf.
Eqs.~(\ref{rerere}) and (\ref{brastable}), respectively.


\subsection{Observables}
\label{sec:observables}
For elastic $\pi N$ scattering, we compare with the dimensionless partial wave amplitudes $\tau$ from
the GWU/SAID analysis~\cite{Arndt:2006bf}. The scattering amplitude $\tau$ for the transition $i\to
f$ in channel space is connected to the amplitude $T$ of Eq. (\ref{bse}) by
\be
\tau_{fi}&=&-\pi\sqrt{\rho_f\,\rho_i}\,T_{fi},\quad\rho=\frac{k\,E\,\omega}{z}
\label{taut}
\ee
where $k\,(E,\omega)$ are the on-shell three-momentum (baryon, meson energies) of the initial or final
meson-baryon system. 

The observables in the reaction $\pi^+p\to K^+\Sigma^+$ can be expressed via the $\tau_{\pi N\to K\Sigma,
I=3/2}$ amplitudes (abbreviated $\tau$ in the following). According to, e.g. Ref.~\cite{Joachain}, for the
scattering of a spin-0 off a spin-$\frac{1}{2}$ particle, the differential cross section for the transition
$(\vec{k}_{i},\nu)\rightarrow(\vec{k}_{f},\nu')$ [$\nu,\,\nu'$ are the $z$-projection of the nucleon spin]
can be expressed as
\cite{Joachain}
\begin{eqnarray}
\frac{d\sigma}{d\Omega}=|\langle s',\nu'|M|s,\nu\rangle|^{2}
\end{eqnarray}
whereas for the initial polarization
\begin{eqnarray}
\vec{P}_{i}=\langle\chi_{\nu}|\vec{\sigma}|\chi_{\nu}\rangle
\end{eqnarray}
and for the final polarization
\begin{eqnarray}
\vec{P}_{f}=\frac{\langle M\chi_{\nu}|\vec{\sigma}|M\chi_{\nu}\rangle}{\langle
 M\chi_{\nu}|M\chi_{\nu}\rangle}.
\end{eqnarray}
$\chi_{\nu}$ is the initial spin-$\frac{1}{2}$ eigenvector and $\vec{\sigma}$ is the Pauli spin-vector.
$M$ can be written in terms of the non-spin-flip and spin-flip amplitudes $g$ and $h$,
\begin{eqnarray}
M=g(k,\theta)\mathds{1}+h(k,\theta)\,\vec{\sigma}\cdot \hat{n}
\end{eqnarray} 
where $g$ and $h$ are complex functions of the energy and scattering angle $\theta$, and
$\hat{n}=\tfrac{\vec{k}_{i}\times \vec{k}_{f}}{|\vec{k}_{i}\times \vec{k}_{f}|}$. The polarization in the
final state $P_{f}$ becomes \cite{Joachain} 
\begin{eqnarray}
\vec{P}_{f}&=&\frac{
(|g|^{2}-|h|^{2})\vec{P}_{i}+(gh^{*}+g^{*}h+2|h|^{2}\vec{P}_{i}\cdot \hat{n})\hat{n}
+i(gh^{*}-g^{*}h)\vec{P_{i}}\times\hat{n} 
}{
|g|^{2}+|h|^{2}+(gh^{*}+g^{*}h)\hat{n}\cdot\vec{P}_{i}}
\end{eqnarray} 
while
\begin{eqnarray}
 \frac{d\sigma}{d\Omega}=\left[\left|g\right|^{2}+\left|h\right|^{2}
  +(g^{*}\,h\,+\,g\,h^{*})\,\hat{n}\cdot\vec{P}_{i}\right]\,\frac{k_{f}}{k_{i}}. 
\end{eqnarray}
In the case of an unpolarized target, $\vec{P}_{i}=0$, one obtains
\begin{eqnarray}
\vec{P}_{f}&=&\frac{(gh^{*}+g^{*}h)}{|g|^{2}+|h|^{2}}\hat{n}= 
 \frac{2\text{Re}(gh^{*})}{|g|^{2}+|h|^{2}}\hat{n}
\end{eqnarray}
and
\begin{eqnarray}
\frac{d\sigma}{d\Omega}&=&(\left|g\right|^{2}+\left|h\right|^{2})\,\frac{k_{f}}{k_{i}}.
\end{eqnarray} 

The $g$ and $h$ amplitudes can be expressed in terms of the partial wave amplitudes according to
\begin{eqnarray}
g&=&\frac{1}{2\sqrt{k_{f}k_{i}}}\non
&\times&\sum_{J}(2J+1) \left( d^{J}_{\frac{1}{2}\frac{1}{2}}(\theta)\left[
 \tau^{J(J-\frac{1}{2})\frac{1}{2}} +
 \tau^{J(J+\frac{1}{2})\frac{1}{2}}\right]\cos\frac{\theta}{2}+ d^{J}_{-\frac{1}{2}\frac{1}{2}}(\theta)
 \left[ \tau^{J(J-\frac{1}{2})\frac{1}{2}}-\tau^{J(J+\frac{1}{2})\frac{1}{2}}\right]
 \sin\frac{\theta}{2}\right)\non
h&=&\frac{-i}{2\sqrt{k_{f}k_{i}}}\non
&\times&\sum_{J}(2J+1) \left( d^{J}_{\frac{1}{2}\frac{1}{2}}(\theta)\left[
 \tau^{J(J-\frac{1}{2})\frac{1}{2}}+\tau^{J(J+\frac{1}{2})\frac{1}{2}}\right]\sin\frac{\theta}{2}-
  d^{J}_{-\frac{1}{2}\frac{1}{2}}(\theta)\left[
 \tau^{J(J-\frac{1}{2})\frac{1}{2}}-\tau^{J(J+\frac{1}{2})\frac{1}{2}}\right]\cos\frac{\theta}{2} \right)
  \ .\non
\non
\end{eqnarray}
The explicit expression for the differential cross section reads
\begin{eqnarray}
\frac{d\sigma}{d\Omega}&=&\frac{1}{2k_{i}^{2}} \frac{1}{2} \,\bigg|
 \sum_{J}(2J+1)(\tau^{J(J-\frac{1}{2})\frac{1}{2}}+
\tau^{J(J+\frac{1}{2})\frac{1}{2}})\cdot d^{J}_{\frac{1}{2}\frac{1}{2}}(\theta)\bigg|^{2}\non
&+&\frac{1}{2k_{i}^{2}} \frac{1}{2}\,
\bigg| \sum_{j}(2J+1)(\tau^{J(J-\frac{1}{2}) \frac{1}{2}}-\tau^{J(J+\frac{1}{2})\frac{1}{2}})
\cdot d^{J}_{-\frac{1}{2}\frac{1}{2}}(\theta)\bigg|^{2}\non
\end{eqnarray}
and the total cross section is obtained by integrating over the solid angle $\Omega$,
\begin{eqnarray}
 \sigma=\frac{1}{2}\cdot \frac{4\pi}{k^{2}_{1}}\sum_{JLS,L'S'} (2J+1)|\tau^{JL'S'}_{LS}|^{2}.
\end{eqnarray}

The spin-rotation parameter $\beta$ is the rotation angle of the spin projection on the scattering plane. It
is given by
\cite{Wolfenstein:1956xg} 
\begin{eqnarray}
\beta=\arctan\left( \frac{2\text{Im}(h^{*}g)}{\left|g\right|^{2}-\left|h\right|^{2}}\right). \label{3.31}
\end{eqnarray}


\section{Results}
\label{sec:1}

%
%

\subsection{Parameters and data base}
\label{sec:fit}
One bare $s$-channel state is included in each of the $I=3/2$ partial waves  S31, P31, D33, D35, F35, F37.
Two are required by data in the P33 wave.  These states were allowed to couple to all $I=3/2$ channels  $\pi
N$, $K\Sigma$, $\pi\Delta$ and $\rho N$. Together with these four bare couplings, the bare mass has to be 
left free as a fit parameter. Thus, there are altogether 40 parameters for the pole part $T^\po$  from
Eq.~(\ref{deco1}). The values of these parameters can be found in Table~\ref{bare_cou}  and the parameter
errors are discussed in Sec.~\ref{sec:uncertainties}. 

The good description of the $\Delta(1232)P_{33}$ resonance shape requires also a fine-tuning of the cut-offs
of the first $s$-channel state in P33, while for all other $s$-channel states, the cut-off was set to 2 GeV
(cf. \ref{sec:bare_res}). Additionally to the $s$-channel parameters, 
the cut-offs of the diagrams of Fig.~\ref{fig:kydiagrams} were adapted 
(results may be found in Table~\ref{tab:cutoff_ky}), while those of the other $t$- and 
$u$-channel diagrams in the model~\cite{Gasparyan:2003fp} were not changed.

We fit to the $\pi^+p\to K^+\Sigma^+$ differential cross section and
polarization, given by the measurements of  Candlin {\it et al.}~\cite{Candlin:1982yv}, available for $z\geq
1822$ MeV. For lower energies we have to resort to the data from Refs.~\cite{Winik:1977mm, Carayannopoulos,
Baltay, Crawford:1962zz, Bellamy:1972fa}. The latter data (see also Ref.~\cite{Livanos:1980vj}) are
compatible with the data of  Ref.~\cite{Candlin:1982yv} in the overlapping energy regions, but usually have
larger errors. The polarization was re-measured later in the higher energy range of the considered
data~\cite{Haba:1988rn}, in consistency with the values of Ref.~\cite{Candlin:1982yv} up to small
deviations. See the captions of Figs.~\ref{fig:diff1} to \ref{fig:pola2} for details. The spin-rotation
parameter $\beta$ for the $\pi^+p\to K^+\Sigma^+$ reaction has been measured in Ref.~\cite{Candlin:1988pn}.
Simultaneously, the polarization has been re-measured in Ref.~\cite{Candlin:1988pn} and consistency with 
results from Ref.~\cite{Candlin:1982yv} was found. In summary, the considered data is consistent and
represents the world data set from threshold to $z=2.35$  GeV. 

For elastic $\pi N$ scattering, the energy-dependent partial wave solution from Ref.~\cite{Arndt:2006bf} up
to $F$ waves is used as input for the fit. Errors have been assigned to it by hand such that the $\pi
N$ data and the $K^+\Sigma^+$ data contribute similarly to the $\chi^2$. 
The uncertainties of the results presented in the following are discussed in Sec.~\ref{sec:uncertainties}.


\subsection{Differential cross section and polarization}

\begin{figure}
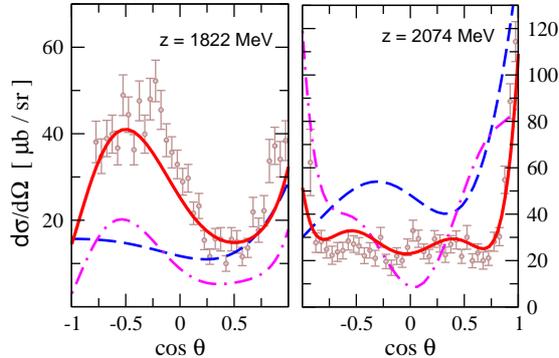

\begin{center}
\includegraphics[height=0.35\textwidth]{dsdo_k+s+_1822_w_pole_part.eps}
\includegraphics[height=0.35\textwidth]{dsdo_k+s+_2074_w_pole_part.eps}
\end{center}
\caption{Contributions to the differential cross section for two typical energies: $T^\npo$ (blue dashed
lines), $T^\po$ (magenta dash-dotted lines), and full solution (red solid lines).}
\label{fig:forward}     
\end{figure}

\begin{figure*}
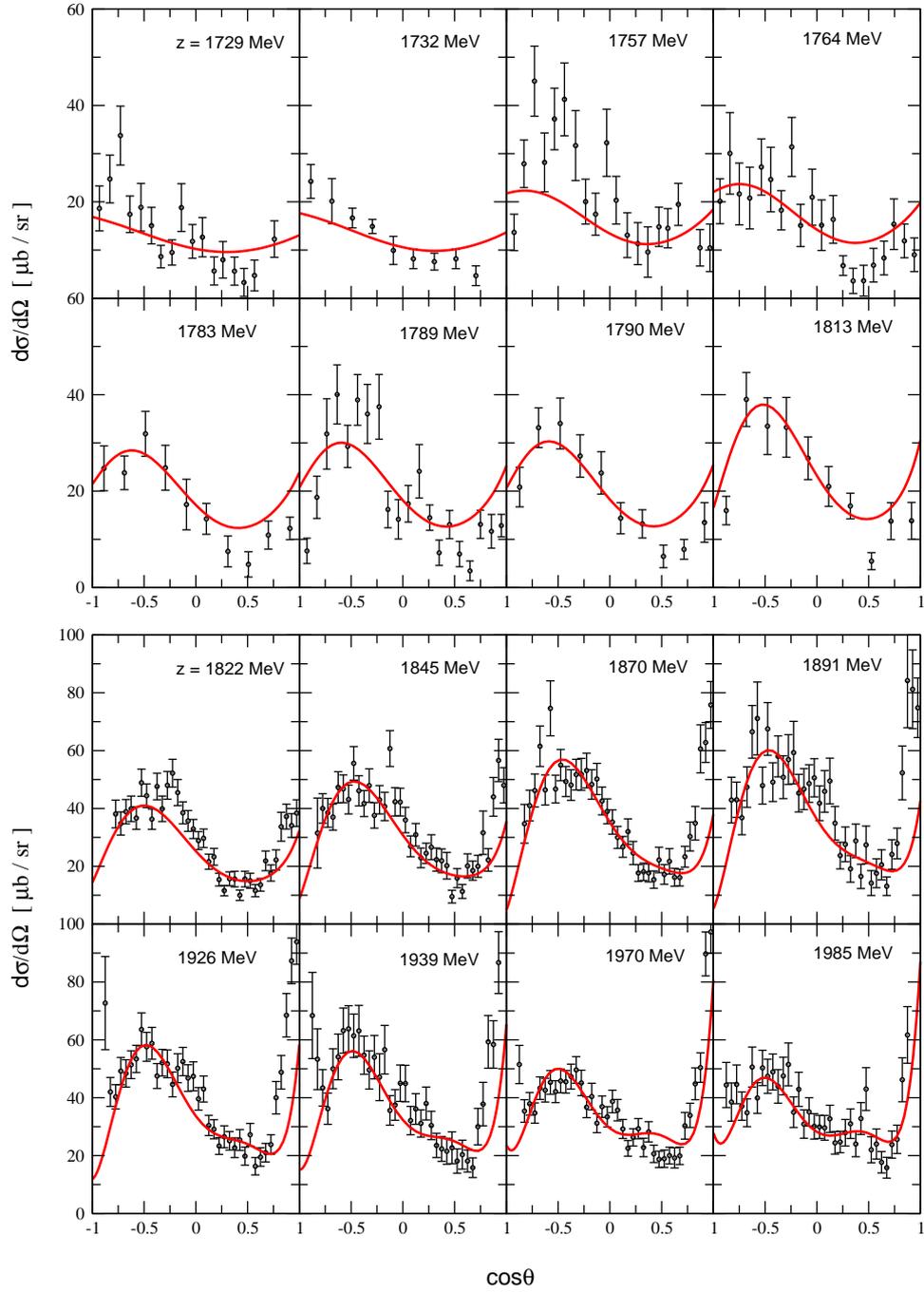

\begin{center}
\includegraphics[width=0.94\textwidth]{dsdo_k+s+_low.eps}\\ 

\vspace*{0.2cm}

\includegraphics[width=0.94\textwidth]{dsdo_k+s+.eps}
\end{center}
\caption{Differential cross section of $\pi^+ p  \rightarrow K^+\Sigma^+$ from $z=1729$ to $z=1985$ MeV.
(Red) solid lines: Present solution. Data: Ref.~\cite{Candlin:1982yv}, except: $z=1729,\,1757,\,1789$ MeV
from Ref.~\cite{Winik:1977mm},  $z=1732,\,1783,\,1813$ MeV from Ref.~\cite{Carayannopoulos}, $z=1790$ MeV
from Ref.~\cite{Baltay}, $z=1764$ MeV from Ref.~\cite{Crawford:1962zz}.}
\label{fig:diff1}      
\end{figure*}

\begin{figure*}
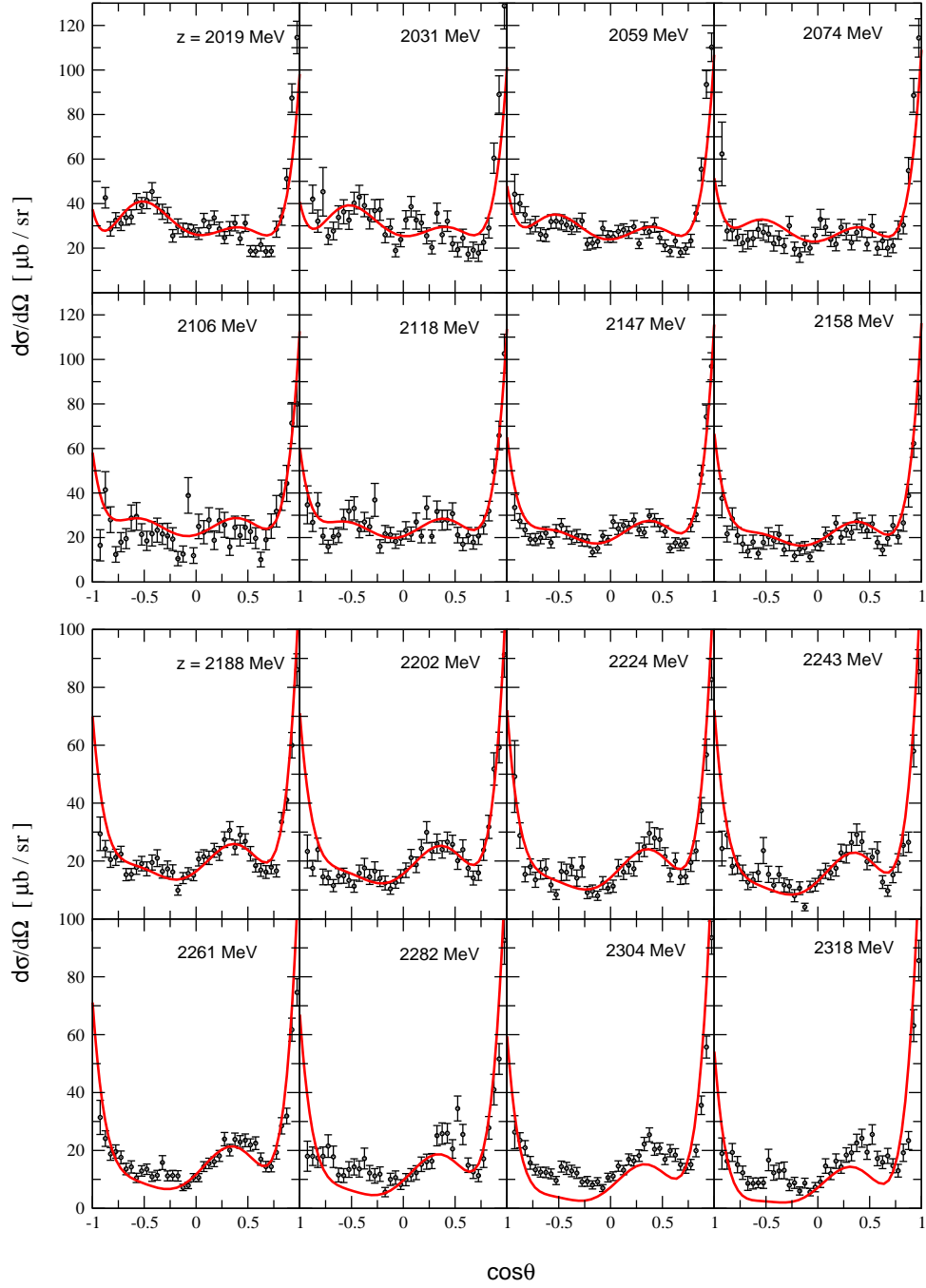

\begin{center}
\includegraphics[width=0.94\textwidth]{dsdo_k+s+_up.eps}\\ 

\vspace*{0.2cm}

\includegraphics[width=0.94\textwidth]{dsdo_k+s+_high.eps}
\end{center}
\caption{Differential cross section of $\pi^+ p  \rightarrow K^+\Sigma^+$ from $z=2019$ to $z=2318$ MeV.
(Red) solid lines: Present solution. Data: Ref.~\cite{Candlin:1982yv}.}
\label{fig:diff2}     
\end{figure*}

The  differential cross sections for the reaction $\pi^+ p  \rightarrow K^+\Sigma^+$ are shown in 
Figs.~\ref{fig:forward}, \ref{fig:diff1} and \ref{fig:diff2}.
The red solid lines show the result of this study. Overall, the data are well described over the entire
energy range. For energies above 2 GeV, we do not claim
validity of the present model, because the analysis of Refs.~\cite{Gasparyan:2003fp, Doring:2009yv} has been
limited to that energy. Consequently, at the highest energies, the $K^+\Sigma^+$ data have not been fitted, 
but up to $z\sim 2.25$ GeV the description of the data is 
still good, as Fig.~\ref{fig:diff2} shows. 

Note, however, that in the present Lagrangian-based framework, the amplitude allows for an extrapolation to
higher energies;  in analyses in which the potential is parameterized purely phenomenologically in terms of
polynomials,  there may be little control on the amplitude outside the fitted energy region.  The overall
agreement seen in Fig.~\ref{fig:diff2} for the higher energies is good although a detailed inspection shows
that there is room for some improvement. In particular, at energies $>$ 2.2 GeV significant deviations are
seen near $\cos~\theta=\pm 0.5$. 

To discuss the individual contributions to the differential cross section,  we show in
Fig.~\ref{fig:forward}, for two typical energies, the non-pole part $T^\npo$ (dashed line), $T^\po$
(dash-dotted line) and the full solution (solid line), see also Eq.~(\ref{deco1}). In forward direction, the
non-pole part $T^\npo$, i.e. the unitarized amplitude from $t$- and $u$-channel exchanges, produces a rise
of the cross sections which becomes more pronounced as the energy increases and which is even stronger than
the experimental forward peak at $z=2074$ MeV.  However, the resonance part $T^\po$ produces a destructive
interference with the $T^\npo$, which is crucial for reproducing the data.  Note especially that $T^\po$ is
a lot more forward-backward symmetric than $T^\npo$ at $z=2074$ MeV.   The forward peak shows the onset of
the $t$-channel dominance which at energies $>$ 3 GeV is most  economically parameterized in terms of Regge
exchanges~\cite{Huang:2008nr,Sibirtsev:2007wk}.

\begin{figure}
\begin{center}
\includegraphics[width=0.8\textwidth]{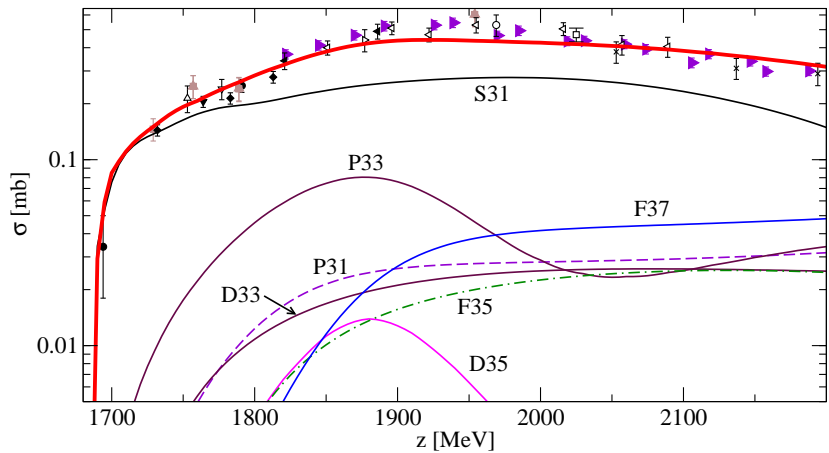} 
\end{center}
\caption{Total cross section of $\pi^+ p \rightarrow K^+\Sigma^+$. (Red) thick solid line: Prediction from
present solution. Also, the contributions from the individual partial
waves are shown, as indicated.
Data: filled triangles right from Ref.~\cite{Candlin:1982yv}, 
other data: see references in the data compilation of Ref.~\cite{Landolt}.}
\label{fig:totalcs}     
\end{figure}

In Fig.~\ref{fig:totalcs}, the total cross section for the reaction $\pi^+ p  \rightarrow K^+\Sigma^+$ is
shown.
The data have not been included in the fit, but the agreement is good. There is a slight underprediction of
$\sigma$ at $z\sim 2$~GeV by the present model, which comes from a slight underprediction of the forward
peak in this energy range, also visible in Figs.~\ref{fig:diff1} and \ref{fig:diff2}. 

\begin{figure}
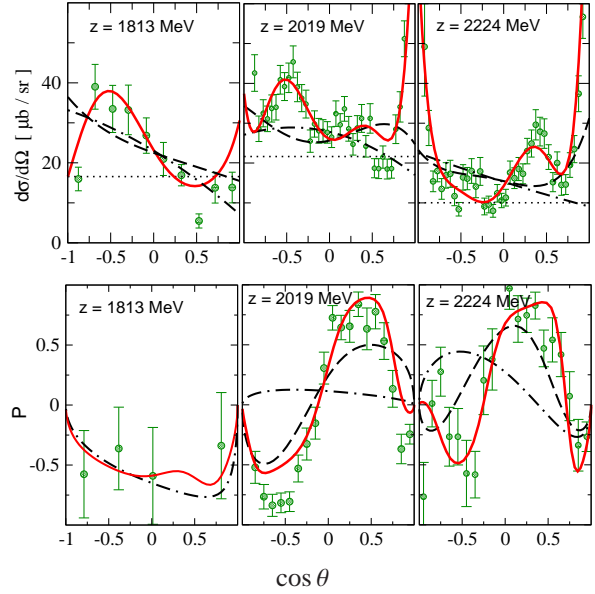

\begin{center}
\includegraphics[height=0.26\textwidth]{dsdo_k+s+_low_spd.eps}  \hspace*{-0.2cm}
\includegraphics[height=0.26\textwidth]{dsdo_k+s+_up_spd.eps}	\hspace*{-0.2cm} 
\includegraphics[height=0.26\textwidth]{dsdo_k+s+_high_spd.eps} \\ 

\vspace*{0.2cm}

\includegraphics[height=0.26\textwidth]{pola_k+s+_low_spd.eps}  \hspace*{-0.2cm}
\includegraphics[height=0.26\textwidth]{pola_k+s+_up_spd.eps}	\hspace*{-0.2cm}
\includegraphics[height=0.26\textwidth]{pola_k+s+_high_spd.eps} 

\vspace*{0.1cm}

$\cos\theta$
\end{center}
\caption{Contribution from different partial waves to differential cross section and polarization for three
typical energies. Dotted lines: $S$ wave. Dash-dotted lines: $S+P$ waves. Dashed lines: $S+P+D$ waves.
Solid lines: $S+P+D+F+G37$ waves (full solution).}
\label{fig:spd}     
\end{figure}

In Fig.~\ref{fig:totalcs}, also the partial cross sections from the individual partial waves are shown.
Except for the S31 and P33 partial waves, all other partial waves are very small; however, for the
differential cross section and polarization, their contributions is essential; indeed, while the removal of
a resonance does not change much the total cross section, the differential observables can change
drastically (see also Fig.~\ref{fig:spd}).

\begin{figure*}
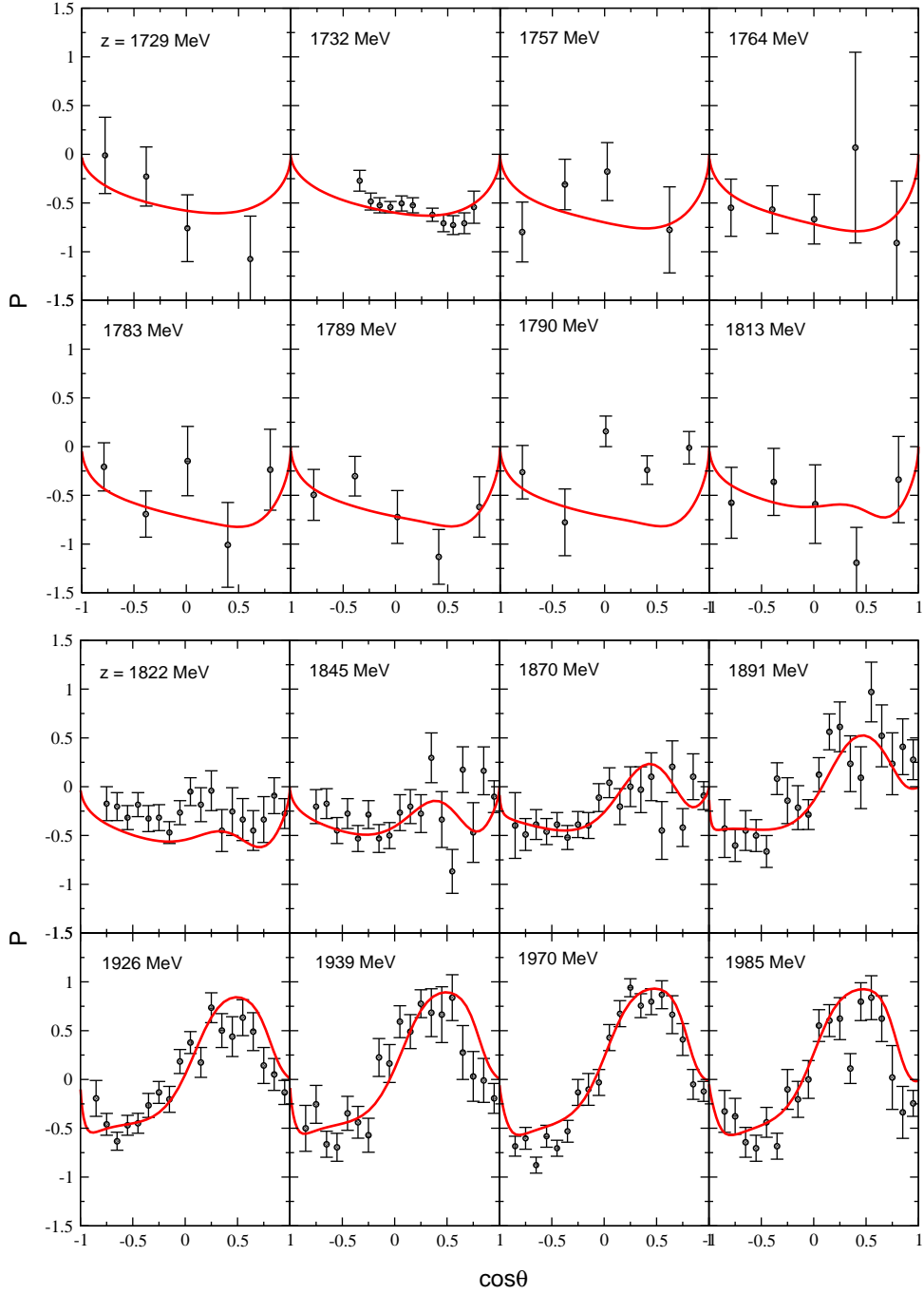

\begin{center}
\includegraphics[width=0.94\textwidth]{pola_k+s+_low.eps}\\ 

\vspace*{0.2cm}

\includegraphics[width=0.94\textwidth]{pola_k+s+.eps}
\end{center}
\caption{Polarization of $\pi^+ p  \rightarrow K^+\Sigma^+$ from $z=1729$ to $z=1985$ MeV.  (Red) solid
lines: Present solution. Data: Ref.~\cite{Candlin:1982yv}, except: $z=1729,\,1757,\,1789$ MeV from
Ref.~\cite{Winik:1977mm},  $z=1782,\,1813$ MeV from Ref.~\cite{Carayannopoulos}, $z=1790$ MeV from
Ref.~\cite{Baltay}, $z=1764$ MeV from Ref.~\cite{Crawford:1962zz}, $z=1732$ MeV from
Ref.~\cite{Bellamy:1972fa}.}
\label{fig:pola1}       
\end{figure*}

\begin{figure*}
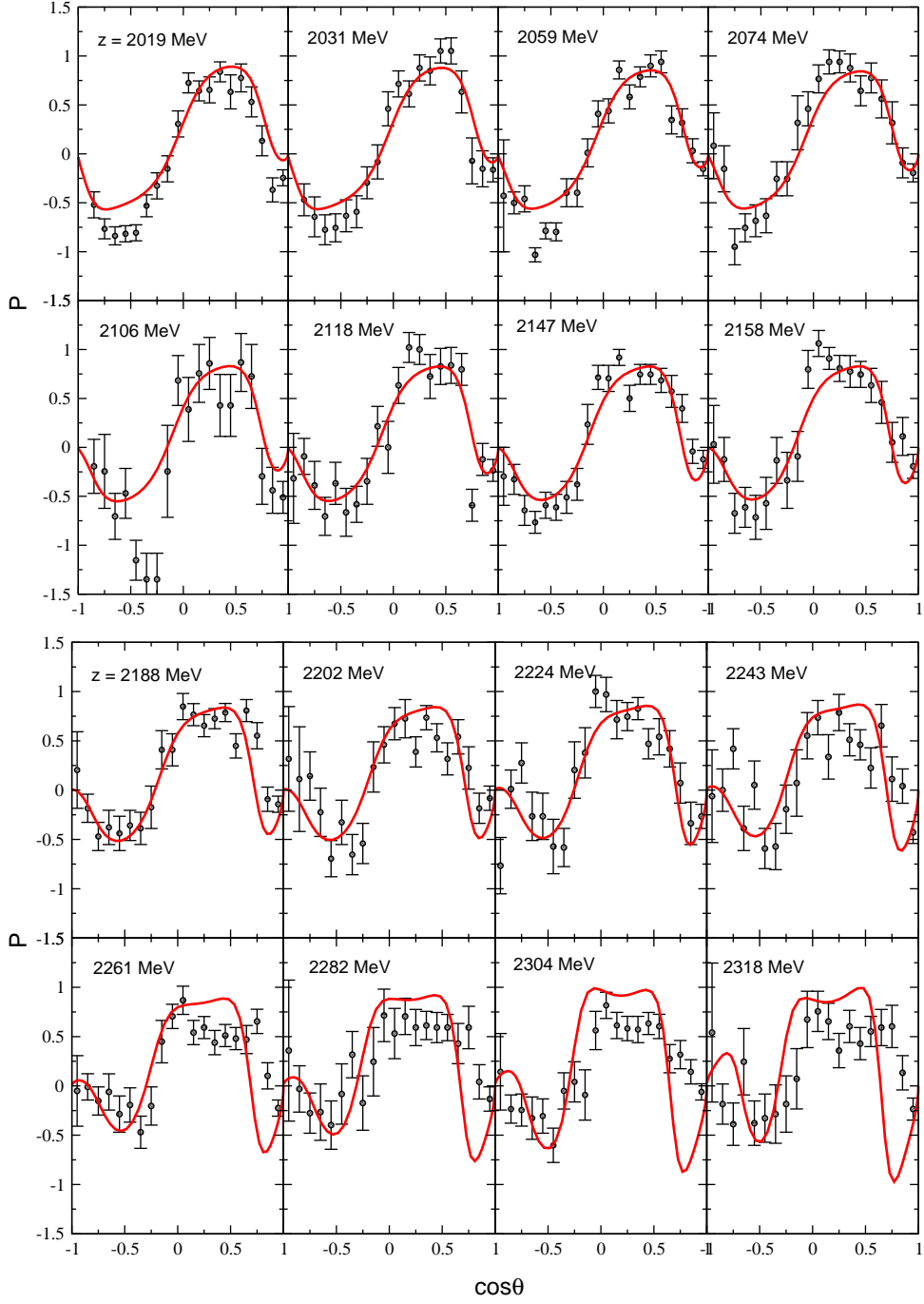

\begin{center}
\includegraphics[width=0.94\textwidth]{pola_k+s+_up.eps}\\ 

\vspace*{0.2cm}

\includegraphics[width=0.94\textwidth]{pola_k+s+_high.eps}
\end{center}
\caption{Polarization of $\pi^+ p  \rightarrow K^+\Sigma^+$ from $z=2019$ to $z=2318$ MeV. Data:
 Ref.~\cite{Candlin:1982yv}. (Red) solid lines: Present solution.}
\label{fig:pola2}      
\end{figure*}

In Figs.~\ref{fig:pola1} and \ref{fig:pola2},
the polarization for the reaction $\pi^+ p  \rightarrow K^+\Sigma^+$ is shown.
Like for the differential cross section, the data show a rich and varying structure over the entire energy
range, and the description by the present model is good. At energies $z>2.2$ GeV, the data have not been
included in the fit and are only plotted for comparison. We found the polarization to be especially
sensitive to the resonance contributions, and the inclusion of these data is important to put constraints on
the corresponding parameters.

The influence of individual partial waves is illustrated in  Fig.~\ref{fig:spd} for three typical energies. 
At lower energies, $S-P$ wave interference is enough to describe the polarization, but not entirely the 
differential cross section (dash-dotted lines at $z=1813$ MeV). For the latter, even a small $F$~wave
admixture is needed to explain the drop at $\cos\theta=-1$ (cf. full solution). For $z=2019$~MeV, one
also needs the $D$ waves for an at least qualitative  description of the polarization (dashed line), and the
$F$~wave is essential in the description of the details of the differential cross section. The same applies
for the highest energy $z=2224$~MeV, where all partial waves are needed to quantitatively describe the data.

\begin{figure}
\begin{center}
\includegraphics[width=0.9\textwidth]{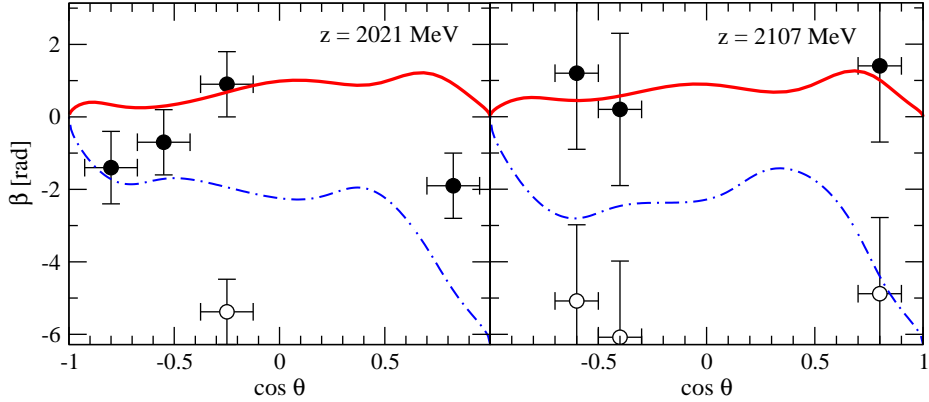} 
\end{center}
\caption{Spin rotation parameter $\beta$ of $\pi^+ p \rightarrow K^+\Sigma^+$ at $z=2021$ and $z=2107$ MeV.
Note that $\beta$ is $2\pi$ cyclic which leads to additional data points at shifted values shown by the
empty circles. Data: Ref.~\cite{Candlin:1988pn}. (Red) solid lines: Prediction from present solution. 
(Blue) dash-dotted lines: Prediction from Ref.~\cite{Candlin:1983cw}.}
\label{fig:spinrot}     
\end{figure}

Fig.~\ref{fig:spinrot}  shows the spin-rotation parameter for the reaction $\pi^+ p  \rightarrow
K^+\Sigma^+$ [cf. Sec.~\ref{sec:observables}].  In this study, $\beta$ is not included in the fit, but
predicted (solid lines). The prediction from the isobar analysis of  Ref.~\cite{Candlin:1983cw} is also
shown (blue dash-dotted lines). 
As $\beta$ is $2\pi$ cyclic, the data from Ref.~\cite{Candlin:1983cw} (solid circles) have been plotted
repeatedly (empty circles).  The present model predicts $\beta$ better than Ref.~\cite{Candlin:1983cw} for
$z=2107$ MeV. Higher precision data would help further pin down the partial wave content because the results
for $\beta$ already show that considering this observable is important to remove ambiguities in the partial
wave content.


\subsection{Partial waves}
\begin{figure}
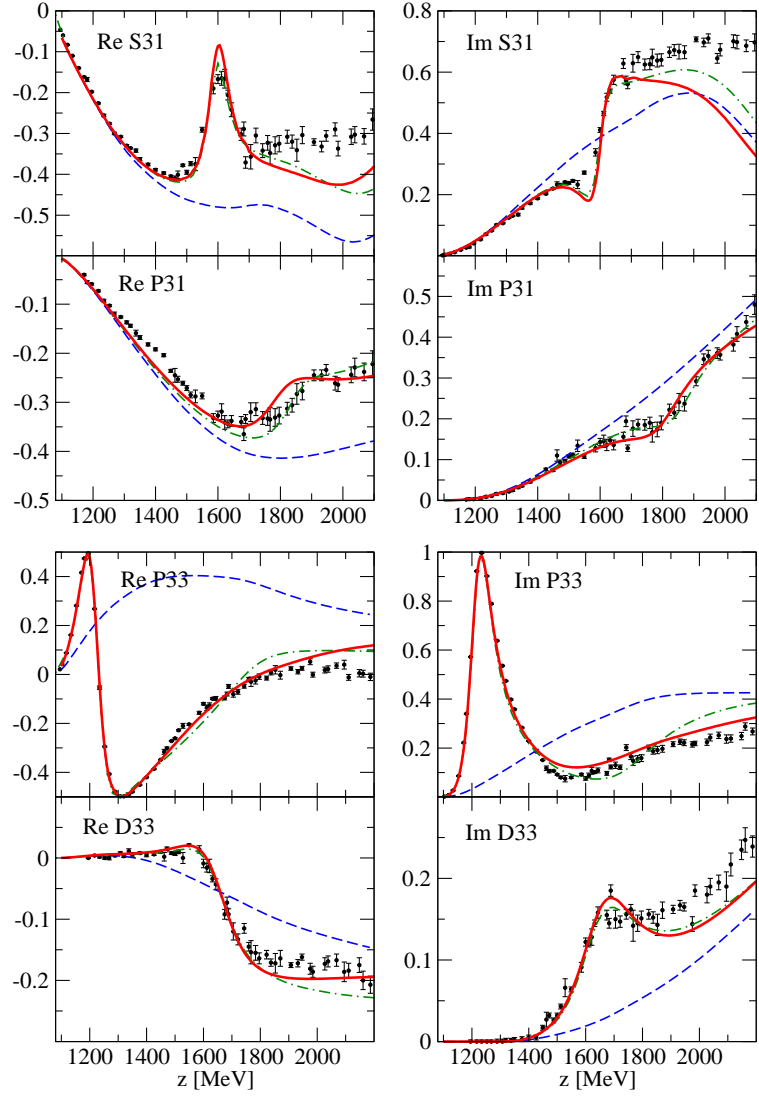
 
\begin{center}
\includegraphics[width=0.73\textwidth]{s31_p31.eps} 

\vspace*{0.2cm}

\includegraphics[width=0.73\textwidth]{p33_d33.eps}
\end{center}
\caption{Elastic $\pi N\to\pi N$ partial waves $S31$, $P31$, $P33$, and $D33$. Data points: GWU/SAID 
partial wave analysis (single energy solution) from Ref.~\cite{Arndt:2006bf}. (Red) solid lines: 
Present solution. (Blue) dashed lines: only $T^\npo$. (Green) dash-dotted lines: J\"ulich model, 
solution 2002 from Ref.~\cite{Gasparyan:2003fp}.}
\label{fig:pw1}       
\end{figure}

\begin{figure}
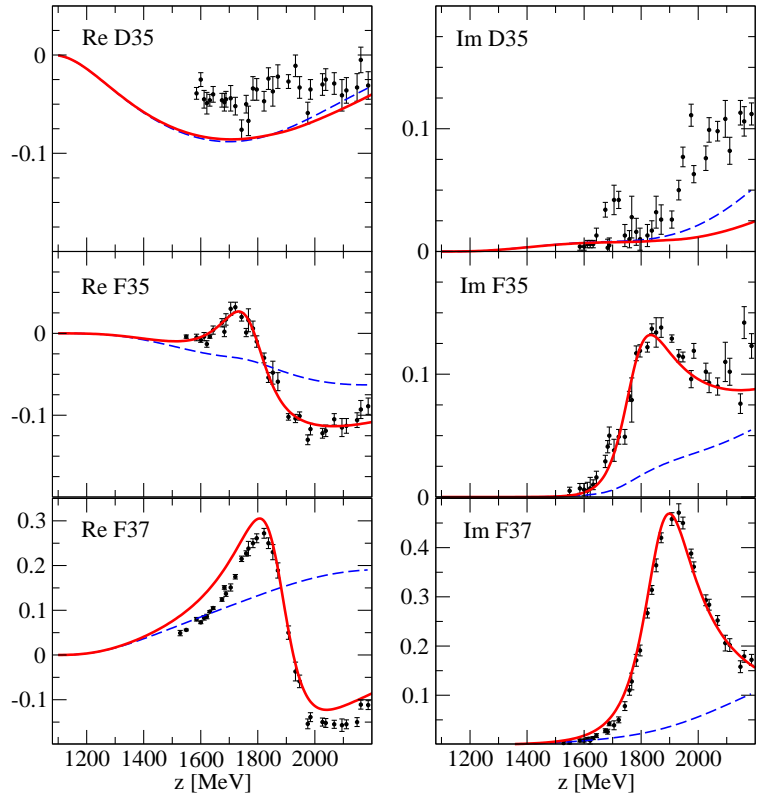

\begin{center}
\includegraphics[width=0.73\textwidth]{d35_f35.eps}\\ 

\vspace*{-0.18cm}

\includegraphics[width=0.73\textwidth]{f37_g17.eps}
\end{center}
\caption{Higher elastic $\pi N\to\pi N$ partial waves $D35$, $F35$, and $F37$. Data points: GWU/SAID 
partial wave analysis (single energy solution) from Ref.~\cite{Arndt:2006bf}. (Red) solid lines: 
Present solution. (Blue) dashed lines: only $T^\npo$.}
\label{fig:pw2}       
\end{figure}

Figs.~\ref{fig:pw1} and \ref{fig:pw2}  show the $I=3/2$ elastic $\pi N\to\pi N$ partial wave amplitudes up
to $J=7/2$, except for $G37$ which is very small.

The result of this study is indicated with the red solid lines.  The data points represent the
energy-independent  partial wave solution from Ref.~\cite{Arndt:2006bf}.  For comparison, also the previous
solution from  2002~\cite{Gasparyan:2003fp} within the framework of the J\"ulich model is shown (green
dash-dotted lines). Note that $J> 3/2$ resonances were not  considered in Ref.~\cite{Gasparyan:2003fp}. The
contribution from $T^\npo$ [cf. Eq.~(\ref{deco1})] is shown with the blue dashed lines. 

The description of the partial waves from the GWU/SAID analysis~\cite{Arndt:2006bf} is comparable to the
results from the  previous J\"ulich analysis~\cite{Gasparyan:2003fp}.   While the present solution is better
for, e.g., the P33 partial wave, some deviations from the GWU/SAID analysis~\cite{Arndt:2006bf} at higher
energies are visible in other partial waves. This may indicate the need for a more systematic fit of the
parameters of $T^\npo$, or may be a sign of the tails of higher lying resonances. Note that similar problems
for the elastic D35 partial wave have been in found in the Gie{\ss}en~\cite{Shklyar:2004dy} analysis, and
also in the EBAC analysis~\cite{JuliaDiaz:2007kz}.

\begin{figure*}
\begin{center}
\includegraphics[width=0.97\textwidth]{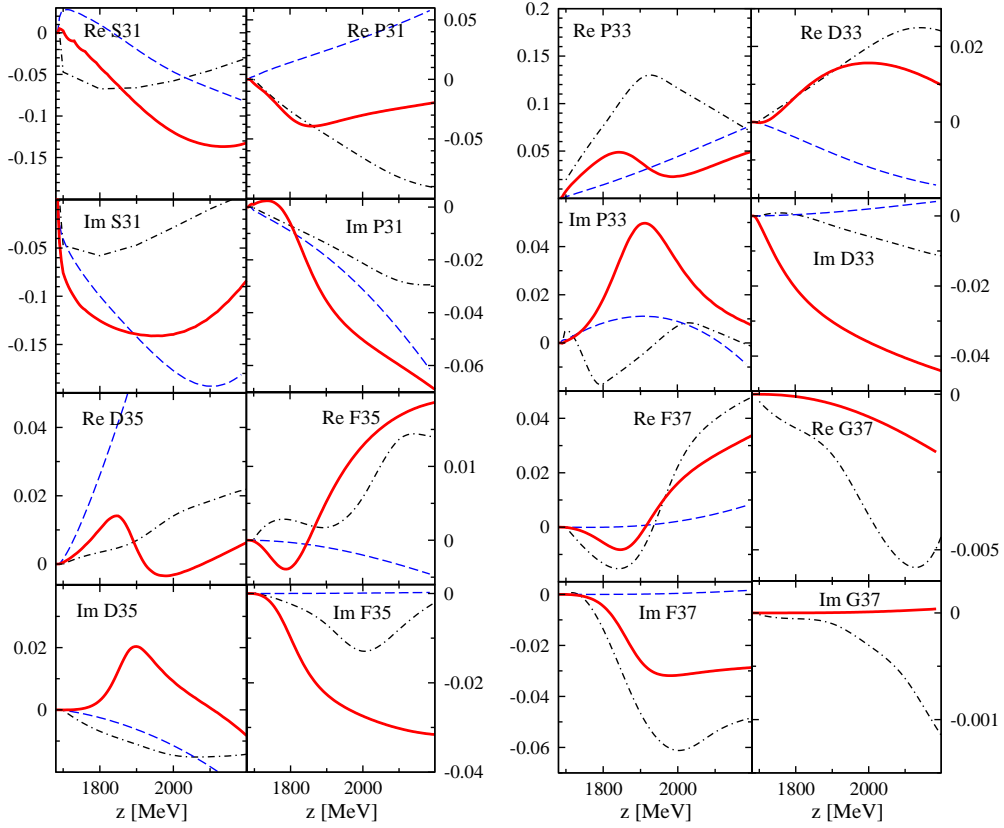} 
\end{center}
\caption{The $\pi^+ p \to K^+\Sigma^+$ partial wave amplitudes. (Red) solid lines: Present solution. 
(Blue) dashed lines: only $T^\npo$. Dash-dotted lines: Analysis from Ref.~\cite{Candlin:1983cw}.}
\label{fig:pwks}     
\end{figure*}

Fig.~\ref{fig:pwks}  shows the $\pi^+ p \to K^+\Sigma^+$ partial wave amplitudes obtained in this study (red
solid lines). 
The contribution from the $t$- and $u$-channel processes, $T^\npo$ from Eq.~(\ref{deco1}), are indicated
with the (blue) dashed lines. The partial wave solution of Ref.~\cite{Candlin:1983cw} is shown with the
dash-dotted lines. Of course, the latter solution cannot be directly compared to the present one, because
there is an overall undetermined phase. Still, even with such a global phase ambiguity, the figure shows
that the partial waves are quite different. In particular, in Ref.~\cite{Candlin:1983cw} lower spin
resonances are only included when providing a substantially improved $\chi^2$ (cf. discussion in
Sec.~\ref{sec:poles}). In the present analysis, we have used all those resonance states up to $J=7/2$ needed
to describe $\pi N$ scattering~\cite{Arndt:2006bf}. Note, at least through coupled channel effects, they
also couple to $K\Sigma$. 


\section{Resonance analysis}


\subsection{Pole positions}
\label{sec:poles}

%
%

\begin{table*}
\caption{Pole positions $z_0$ of resonances with isospin $I=3/2$. For each resonance, the upper row shows
 Re~$z_0$~[MeV], the lower -2~Im~$z_0$~[MeV]. The results of the present calculation are shown in the column
 {\sl J\"ulich}. The Table also specifies the type of analysis: Dynamical coupled-channels model [DCM],
 $K$-matrix approach [KM], dispersion analysis [DA], or isobar analysis [IA]. The quoted values are either
 pole positions [P], Breit-Wigner value [BW], or pole positions obtained by speed plot techniques
 [SP]. For the other entries, see text. Uncertainty in the last digits in parentheses, (a) indicates that 
 the corresponding resonance is dynamically generated in the present approach.}
\begin{center}
\begin{footnotesize} 
\renewcommand{\arraystretch}{1.33}
\begin {tabular}{cc|c|ccc|cc|c|c|cc} 
\hline\hline
{\bf Data:}&\multicolumn{2}{c|}{$\pi N$ + $K^{+}\Sigma^{+}$ ($+\cdots$)}&\multicolumn{5}{c|}{$\pi N$} & $K^{+}\Sigma^{+}$	& $\pi\pi N$	&\multicolumn{2}{c}{Quark Models} 		              \\
{\bf Analysis: }        & J\"ulich 	&Gie{\ss}en	& GWU		&KH	&CMB	&EBAC	&DMT	& Cdl 	& Mnly 		& LMP, ${\cal A}$& CI  	      	\\
{\bf Type: }		& DCM		& KM		& KM/DA		&DA	&DA	&DCM	&DCM	& IA	& KM		& ---		& ---		\\
{\bf Pole/BW:}		& P		& BW		& P		&SP	&P	& P	& P	& BW	& BW		& ---		& ---         	\\
\hline
 $\Delta(1232)P_{33}$ 	& 1216      	& 1228(1)	& 1211 		&1209	&1210	& 1211	& 1212	& ---   &  1232		& 1261 	 	& 1230	      	\\
 3/2$^{+} $ ****   	& 96      	& 106(1)	& 99  		&100	&100	& 100	& 98	& 	&   118 	&  ---  	& ---           \\
\hline
 $\Delta(1600)P_{33}$ 	& 1455$^{(a)}$ 	& 1667(1)	& 1457 		&1550	&1550	& ---	& 1544	& ---   &  1706		& 1810		& 1795        	\\
 3/2$^+$ ***       	& 694       	& 397(10)	& 400 		& ---	&200	&	& 190	& 	&   430 	&  ---  	& ---         	\\
\hline
 $\Delta(1620)S_{31}$ 	& 1599      	& 1612(2)	& 1595 		&1608	&1600	& 1563	& 1589	& ---   &   1672 	& 1654 		& 1555        	\\
1/2$^-$ ****       	& 62        	& 202(7)	&  135 		&116	&120	& 190	& 148	& 	&   154  	&  ---   	& ---         	\\
\hline
 $\Delta(1700)D_{33}$ 	& 1644      	& 1678(1)	& 1632 		&1651	&1675	& 1604	& 1604	& ---  	& 1762 		& 1628 		& 1620        	\\
3/2$^-$ ****       	& 252       	& 606(15)	&  253 		&159	&220	& 212	& 142	&   	&   599 	&  ---  	& ---         	\\
\hline
$K^+\Sigma^+$(1688)	&           	&		&      		&	&	&	&	&     	& 		&      		&             	\\
\hline
$ \Delta(1750)P_{31}$ 	& 1668$^{(a)}$	& 1712(1)	& 1771 		& ---	& ---	& ---	& ---	& ---  	& 1744 		& 1866 		& ---        	\\
1/2$^+$ *          	&  892      	& 643(17)	&  479 		& 	& 	& 	& 	&   	& 299  		& ---   	&          	\\
\hline
 $\Delta(1900)S_{31}$ 	&  ---		& 1984		& ---  		&1780	&1870	& ---	& 1774	& ---  	&  1920 	& 2100		& 2035         	\\ 
1/2$^-$ **         	&       	& 237		&   		&170	&180	& 	& 72	&    	&   263   	& ---   	& ---          	\\
\hline
$ \Delta(1905)F_{35}$ 	& 1764      	& 1845(15)	& 1819 		&1829	&1830	& 1738	& 1760	&  1960 & 1881 		& 1897		& 1910	      	\\
5/2$^+$ ****       	&  218      	& 426(26)	&  247 		&303	&280	& 220	& 200	&   270 &  327 		&  ---		& ---         	\\
\hline
$ \Delta(1910)P_{31}$ 	& 1721      	& 1975		& 1771 		&1874	&1880	& ---	& 1900	& ---	& 1882	   	& 1906	   	& 1875	 	\\
1/2$^+$ ****       	&  323      	& 676		&  479 		&283	&200 	& 	& 174	&   	& 239	   	& ---	   	& ---	 	\\
\hline
$ \Delta(1920)P_{33}$ 	& 1884      	& 2057(1)	& ---    	&1900	&1900	& ---	& ---	& 1840 	& 2014 		& 1871		& 1915        	\\  
3/2$^+$ ***        	&  229      	& 525(32)	& 	    	& ---	&300	& 	&  300  &  200 	&  152   	&  ---  	& ---	      	\\
\hline
$ \Delta(1930)D_{35}$ 	& 1865      	& ---		& 2001		&1850	&1890	& ---	& 1989	& --- 	& 1956	   	& 2179	   	& 2155	 	\\
5/2$^-$ ***        	&  147      	& 		&  387 		&180	&260 	& 	& 280	&   	&  526	   	&  ---	   	& ---	 	\\
\hline
 $\Delta(1940)D_{33}$ 	& ---		& ---		& ---		& ---	& ---	& ---	& ---	& ---	& 2057		& 2089		& 2080	      	\\
3/2$^-$ *          	& 		& 		& 		& 	& 	& 	& 	& 	&  460  	& ---		& ---	      	\\
\hline
$ \Delta(1950)F_{37}$ 	& 1873      	& ---		& 1876 		&1878	&1890	& 1858	& 1858	& 1925 	& 1945  	& 1956		& 1940	      	\\ 
7/2$^+$ ****       	&  206      	& 		&  227 		&230	&260	& 200	& 208	&  330 	&  300   	& ---  		& ---         	\\
\hline
\hline
\end{tabular}
\end{footnotesize}
\end{center}
\label{tab:popo}
\end{table*}

In Table~\ref{tab:popo}, the pole positions found in the present analysis are shown as {\it J\"ulich}. The
positions are visualized in Fig.~\ref{fig:plane} together with the chosen directions of the branch cuts (cf.
Sec.~\ref{sec:analytisch}).
The first line of Table~\ref{tab:popo} indicates the data that have been taken into account in the different
analyses, the second to fourth lines indicate the analyses (see below), their type, and whether the quoted
values are pole positions or Breit-Wigner parameters.

As for the well-established 4-star resonances, it is no surprise that the pole positions found in this study
are in agreement with the values from the GWU/SAID analysis~\cite{Arndt:2006bf} (4th column), because the
partial waves from that analysis serve as input for the present study. Indeed, the $\Delta(1232)P_{33}$,
$\Delta(1700)D_{33}$, $\Delta(1905)F_{35}$, and $\Delta(1950)F_{37}$ show clear signals in $\pi N\to\pi N$
(cf. Figs.~\ref{fig:pw1} and \ref{fig:pw2}) and the present fit agrees well with these partial waves. The
$\Delta(1620)S_{31}$ is narrower than in the GWU/SAID analysis which comes from the weight of the $K\Sigma$
data in the present fit. Also for the $P_{31}(1910)$ resonance, the width is different from the  one found
in GWU/SAID; this may come from the small resonance signal in elastic $\pi N$ scattering on top of a large
background (cf. Fig.~\ref{fig:pw1}). However, note that the present formalism is not a $K$-matrix approach;
due to the dispersive parts present in this analysis, in principle one cannot expect similar pole positions,
even if the amplitudes are very similar on the real, physical axis. 

The $\Delta(1920)P_{33}$ and $\Delta(1930)D_{35}$ resonance show no or very small resonance signals in the
GWU/SAID analysis of elastic $\pi N$ scattering (cf. Table~\ref{tab:bra1}). Their position is, thus, barely
fixed from elastic $\pi N$ scattering.  It is then interesting to note that the constraints from the
$K^+\Sigma^+$ data  lead to resonance positions in vicinity to those quoted in the PDG, rated with 3 stars.
Thus, we can accumulate further evidence for these states and their positions. It should be stressed again
that the resonance positions are not preassigned in the present ansatz, but left completely free in the
fit.

Finally, we find poles in the scattering amplitude which are not induced by bare $s$-channel resonance
states. Those poles are already present in $T^\npo$ [cf. Eq.~(\ref{deco1})] and arise from the unitarization
of the $t$- and $u-$channel exchange diagrams. These dynamically generated poles are, in the present
analysis, far in the complex plane: a $\Delta(1600)P_{33}$ and a $\Delta(1750)P_{31}$. Apart from these two
poles listed in Table~\ref{tab:popo}, we find a very wide dynamically generated pole in  the S31 partial
wave at $z_0=2170-645\,i$ MeV and one in the D35 partial wave at $z_0=2734-445\,i$ MeV
which thus have widths $\Gamma=-2\,{\rm Im}\,z_0$ of around 1 GeV. Both these states
are too wide to be identified with resonances quoted  in the PDG~\cite{pdg}. 
This applies also to the
$P_{31}$ state, while for the $\Delta(1600)P_{33}$  state quoted in Table~\ref{tab:popo} there may also be
some evidence in the GWU/SAID analysis for a wide state. 

Note that not all those states included here by  bare $s$-channel diagrams are necessarily genuine
resonances; once the channel space is enlarged appropriately by inclusion of, e.g.,  $K\Sigma^*$, resonances
like the $\Delta(1700)D_{33}$ may appear dynamically generated. This is discussed in 
Ref.~\cite{Doring:2010fw} where the prediction  of the $I^S$ and $I^C$ observables in the reaction $\gamma p
\to\pi^0\eta p$ is shown to coincide well with experiment suggesting a dynamical nature for that resonance.

The column of Table~\ref{tab:popo} marked {\it Gie{\ss}en} shows results of the $K$-matrix based analysis
from the Gie{\ss}en  group~\cite{Penner:2002ma,Shklyar:2004dy}, see also Introduction. While  resonances
with spin $5/2$ have been  included recently~\cite{Shklyar:2004dy}, the $\Delta(1950)F_{37}$ resonance is
not, which plays an important role in $K^+\Sigma^+$ production~\cite{Candlin:1983cw}. 
Also, the absence of
some analytic properties restricts the model to real energies, and thus no pole positions can be quoted. The
numbers shown in Table~\ref{tab:popo} are, thus, Breit-Wigner parameters.

Table~\ref{tab:popo} shows also the pole positions from the three standard partial wave analyses of elastic
$\pi N$ scattering, marked as {\it GWU} (George Washington University)~\cite{Arndt:2006bf}, 
{\it KH} (Karlsruhe-Helsinki) \cite{Hoehler1, Hoehler2}, and {\it
CMB} (Carnegie-Mellon-Berkeley)~\cite{Cutkosky}. 

The following column shows the pole positions from the extraction of the EBAC group as quoted in
Refs.~\cite{Suzuki:2008rp,Suzuki:2009nj}, based on the analysis of elastic $\pi N$ scattering of 
Ref.~\cite{JuliaDiaz:2007kz}. For a review on the theoretical foundations of the formalism, see
Ref.~\cite{Matsuyama:2006rp}. The framework has many similarities to the present one, although there are
differences such as the treatment of the nucleon pole or the role of the Roper resonance, which appears
dynamically generated in the present framework~\cite{Krehl:1999km} but is included as a genuine state in the
EBAC model~\cite{Suzuki:2009nj}. 

The column {\it DMT} shows the recent pole extraction~\cite{Tiator:2010rp} from the Dubna-Mainz-Taipeh
analysis~\cite{Chen:2007cy} of elastic $\pi N$ scattering. Like the EBAC and the present model, this
approach is a dynamical meson exchange model, i.e., not a $K$-matrix approach. 

The column {\it Cdl} shows the Breit-Wigner parameters obtained from the isobar analysis of
Ref.~\cite{Candlin:1983cw}. In the isobar analysis, a purely phenomenological background and resonances are
added in a way violating unitarity, and the fit is exclusively to the $\pi^+p\to K^+\Sigma^+$ data of
Ref.~\cite{Candlin:1982yv}. The spin-rotation parameter $\beta$ has been evaluated in 
Ref.~\cite{Candlin:1988pn} using this analysis resulting in poor agreement with the data (cf. discussion of
Fig.~\ref{fig:spd}). Relatively few resonances are quoted in Table~\ref{tab:popo}, because in the
isobar analysis~\cite{Candlin:1983cw} only those lower-spin resonances are considered that lead to a
significant improvement of the $\chi^2$.

The column {\it Mnly} shows the results of the $K$-matrix analysis of the $\pi
N\to\pi\pi N$ reaction of Ref.~\cite{Manley:1992yb}.

The pole positions from the multi-channel CMB type analysis of Vrana, Dytman, and Lee of
Ref.~\cite{Vrana:1999nt} are not shown in Table~\ref{tab:popo}. This analysis finds two S31 resonances, two
P31 and three P33 resonances, one D33 and one F37 resonances. It does not find the second D33 state of
Table~\ref{tab:popo}, but two D35 and two F35 resonances in contrast to the resonances quoted in
Table~\ref{tab:popo}.

The last two columns show some of the predictions from the quark models of L\"oring, Metsch, 
Petry~\cite{Loring:2001kx} and Capstick, Isgur~\cite{Capstick:1986bm}.  In the work of
Ref.~\cite{Loring:2001kx}, the non-strange constituent quark mass and two confinement parameters are
fitted to the $\Delta$-Regge trajectory.  All mass values in Table~\ref{tab:popo} except the
$\Delta(1232)P_{33}$ and the $\Delta(1950)F_{37}$ are then predictions. 

The question arises to which extent the poles found in this analysis can be related to the quark
model states quoted in Table~\ref{tab:popo}~\cite{Loring:2001kx, Capstick:1986bm} or
others~\cite{Isgur:1978xj,Capstick:1992th,Capstick:1993kb}. As Table~\ref{tab:popo} shows,  the quark models
predict the baryonic resonance spectrum quite well. However, the hadronic dressing effects are not explicit
in these calculations, and they can be large. In fact, we have found large correlations between bare masses,
coupling constants, the used channel space and the renormalization scheme~\cite{Doring:2009bi}, and the
matching point between quark models on one side and dynamical coupled-channels approaches on the other side
is still an open issue. 


\subsection{Branching ratios}
\label{sec:branching}

%
%

\begin{table}
\caption{Left: $\pi N\to\pi N$ residues of the present study ({\it J\"u}) and from Ref.~\cite{Arndt:2006bf}
({\it GWU}). For each resonance, the upper row shows $|r|$~[MeV],  the lower $\theta$~[$^0$]. Right: $\pi N$
branching ratios in \%. (a) indicates that 
 the corresponding resonance is dynamically generated in the present approach.}
\begin{center}
\renewcommand{\arraystretch}{1.30}
   \begin {tabular}{c|cc|cc} \hline\hline
\multicolumn{5}{c}{$\pi N\to\pi N$}\\  \hline
&\multicolumn{2}{c|}{$|r|,\theta$}&\multicolumn{2}{c}{$\Gamma_{\pi N}/\Gamma_{\rm tot}$} \\
			& J\"u	& GWU	&J\"u	&GWU \\
\hline
 $\Delta(1232)P_{33}$ 	& 49.3	& 52	&100	& 100  \\
 3/2$^{+} $ ****   	& -40.5	& -47   &	&      \\
\hline
 $\Delta(1600)P_{33}$ 	& 101	& 44	&24	& ---  \\
 3/2$^+$ *** $^{(a)}$  	& -196	& +147  &	&      \\
\hline
 $\Delta(1620)S_{31}$ 	& 14	& 15	&47	& 32   \\
1/2$^-$ ****       	& -107	& -92   &	&      \\
\hline
 $\Delta(1700)D_{33}$ 	& 21	& 18	&16	& 16   \\
3/2$^-$ ****       	& -40	& -40   &	&      \\
\hline
$K^+\Sigma^+$(1688)	&	&	&	&      \\
\hline
$ \Delta(1750)P_{31}$ 	& 18	& ---   &3.4	& ---  \\
1/2$^+$ * $^{(a)}$    	& -300	& ---   &	&      \\
\hline
$ \Delta(1905)F_{35}$ 	& 11	& 15	&10	& 12   \\
5/2$^+$ ****       	& -45	& -30	&	&      \\
\hline
$ \Delta(1910)P_{31}$ 	& 13	& 45	&8.1	& 24   \\
1/2$^+$ ****       	& -175	& +172	&	&      \\
\hline
$ \Delta(1920)P_{33}$ 	& $<$1	& ---	&$<$1	& ---  \\
3/2$^+$ ***        	& -114	& ---	&	&      \\
\hline
$ \Delta(1930)D_{35}$ 	& $<$1	& 7	&$<$1	& 8    \\
5/2$^-$ ***        	& -358	& -12	&	&      \\
\hline
$ \Delta(1950)F_{37}$ 	& 47	& 53	&45	& 47   \\
7/2$^+$ ****       	& -30	& -31	&	&      \\
\hline
\hline
   \end {tabular}
\end{center}
\label{tab:bra1}
\end{table}

In Table~\ref{tab:bra1} the residues and resulting branching ratios into the $\pi N$ channel are shown as
obtained in the present study. The values are compared to the GWU/SAID results~\cite{Arndt:2006bf} [note
that the values of $\Gamma_{\pi N}/\Gamma_{\rm tot}$ from Ref.~\cite{Arndt:2006bf} have been obtained from a
Breit-Wigner fit, while the values of the present study are directly obtained from the residues using
Eq.~(\ref{brastable})]. 

As already noted in the discussion of the pole positions, most of the residue strengths and phases of the
present analysis coincide quite well with those of the GWU/SAID analysis. As Table~\ref{tab:bra1} shows,
this is the case for the 4-star resonances $\Delta(1232)P_{33}$, $\Delta(1620)S_{31}$,
$\Delta(1700)D_{33}$, $\Delta(1905)F_{35}$,  and $\Delta(1950)F_{37}$. For resonances that couple weakly to
the $\pi N$ channel or which are very wide, such as the $\Delta(1910)P_{31}$, the differences are larger.
The dynamically generated
$\Delta(1750)P_{31}$ quoted in Table~\ref{tab:popo} is too wide to be considered a resonance state. 

However, the
dynamically generated $\Delta(1600)P_{33}$ resonance is also seen in the GWU/SAID
analysis~\cite{Arndt:2006bf}, with qualitatively similar properties  (very wide, rather small to medium
branching ratio into $\pi N$, similar residue phase $\theta$). It should be stressed that the appearance of
dynamically generated states in the present model is strongly restricted  by the fact that the generating
$t$- and $u$-channel processes connect all partial waves at the same time; there is little or no room to
manipulate the strengths of these transitions in order to generate poles, without immediate consequences for
all other partial waves. Thus, dynamically generated poles are rather stable objects --- all of those
found in this study are already present in the
solution from 2002 of the J\"ulich model~\cite{Gasparyan:2003fp}, although they have not been searched for
because the analytic continuation became available only in Ref.~\cite{Doring:2009yv}. 

The attraction that leads to the dynamical generation of the $\Delta(1600)P_{33}$ and the
$\Delta(1750)P_{31}$ comes mainly from the $\pi\Delta$ channel: The coupling of the $\Delta(1600)P_{33}$
into the $(\pi\Delta)_{\rm P33}$ channel [cf. Eq. (\ref{couplings})] has a modulus of $|g|=17\cdot 10^{-3}\,
{\rm MeV}^{-1/2}$ [$\Delta(1750)P_{31}\to (\pi\Delta)_{\rm P31}:$ $|g|=20$] which is of the same size as the
coupling of the $\Delta(1232)P_{33}$ to $\pi N$ (19 in these units).  Indeed, the dominant decay
channel of the $\Delta(1600)P_{33}$, quoted by the PDG~\cite{pdg}, is the one to $\pi\Delta$ (40-70\%).
Electromagnetic probes could shed further light on the nature of this resonance, as recently discussed
using hadronic dressing and a constituent quark model~\cite{Ramalho:2010cw}.

The $\Delta(1930)D_{35}$ couples extremely weakly to the $\pi N$ channel in this analysis. Interestingly,
the GWU/SAID analysis also finds a resonance in $D35$ with a very small signal in $\pi N$ scattering [cf.
Table~\ref{tab:bra1}]. Also the $\Delta(1920)P_{33}$ resonance has a very small $\pi N$ branching ratio in
the present study --- note that there is no pole found in the GWU/SAID analysis of elastic $\pi N$
scattering. However, removing one or both of these states in the present analysis,  the
$K^+\Sigma^+$ data are described much worse, even if all other resonance parameters are refitted.

\begin{table}
\caption{Left: $\pi^+ p\to K^+\Sigma^+$ residues $|r|$~[MeV], $\theta$~[$^0$] of the present study. Right: 
Transition branching ratio [\%] in the present study ({\it J\"u}), 
from Ref.~\cite{Candlin:1983cw} ({\it Cdl}), and from Refs.~\cite{Penner:2002ma,Shklyar:2004dy} ({\it 
Gie{\ss}}). Uncertainty in the last digit in parentheses.}
\begin{center}
\renewcommand{\arraystretch}{1.30}
   \begin {tabular}{c|c|ccc} \hline\hline
\multicolumn{5}{c}{$\pi^+ p\to K^+\Sigma^+$}\\  \hline
&$|r|,\theta$&\multicolumn{3}{c}{$(\Gamma^{1/2}_{\pi N}\Gamma^{1/2}_{K\Sigma})/\Gamma_{\rm tot}$}\\ 
&J\"u&J\"u&Cdl&Gie{\ss}\\
\hline
$ \Delta(1905)F_{35}$  &1.4    & 1.23  & 1.5(3)	& $<$1  \\
5/2$^+$ ****           & -313  &       &	&   	\\
\hline
$ \Delta(1910)P_{31}$  & 5.5   & 2.98  & $<$3	& 1.1   \\
1/2$^+$ ****           & -6    &       &	&   	\\
\hline
$ \Delta(1920)P_{33}$  & 5.9   & 5.07  & 5.2(2)	& 2.1(3)\\
3/2$^+$ ***            & -38   &       &	&   	\\
\hline
$ \Delta(1930)D_{35}$  & 1.6   & 2.14  & $<$1.5	&       \\
5/2$^-$ ***            & -43   &       &	&	\\
\hline
$ \Delta(1950)F_{37}$  & 2.7   & 2.54  & 5.3(5)	& ---   \\
7/2$^+$ ****           & -255  &       &	&  	\\
\hline
\hline
   \end {tabular}
\end{center}
\label{tab:bra2}
\end{table}

In the second column of Table~\ref{tab:bra2}, the present results for the residues into the $K\Sigma$
channel are shown. On the right-hand side, transition branching ratios of $\pi N\to K\Sigma$ from different
analyses are displayed. Again, {\it J\"u} marks the present results. 

The values for $(\Gamma^{1/2}_{\pi N}\Gamma^{1/2}_{K\Sigma})/\Gamma_{\rm tot}$ are quite different from each
other. The only common feature is the prominent role of the $\Delta(1920)P_{33}$ resonance. Also, the $F37$
resonance has been found important in the present analysis and in Ref.~\cite{Candlin:1983cw}. 

This wave is missing in the Gie{\ss}en analysis which may distort the resonance content and branching ratios
to $K\Sigma$ and may explain the differences observed in Table~\ref{tab:bra2}. As for Candlin's
analysis~\cite{Candlin:1983cw}, there are the above-mentioned conceptual problems of the isobar analysis, in
particular the oversimplified and unitarity violating construction of the partial wave amplitudes;
discrepancies to the present results are, thus, expected.

There are also older analyses, not quoted in the Table~\cite{Livanos:1980vj, Deans:1974yv}, which are based
on low-statistic data previous to the one published in Ref.~\cite{Candlin:1982yv}. The branching ratios
obtained in these analyses show large discrepancies among each other and also to those quoted in
Table~\ref{tab:bra2}.

All the discussed analyses differ in the data bases considered, the theoretical tools used, and
third, by the quality of the fits. Moreover, the partial wave content is not unique even within the same
framework and even if data from different reactions are combined into a global fit --- for a clear
demonstration of this, see Ref.~\cite{Nakayama:2008tg}. Still, in the present approach, the explicit
microscopical treatment of the non-pole part provides a realistic background which helps minimize
ambiguities from resonance contributions. Furthermore, the joint treatment of elastic $\pi N$ and
$K^+\Sigma^+$ data helps determine more precisely the resonance content of the $K^+\Sigma^+$ production
amplitude. 


\section{Uncertainties}
\label{sec:uncertainties}
In this section we give some remarks on the reliability of the resonance parameters extracted based on the
input data used --- we will make no attempt to estimate the theoretical uncertainty of the approach as such.
We are not (yet) able to quantify the uncertainty introduced into the analysis by the particular formalism
used. In principle, once a set of model analyses exists fitted to the same data with the same channels
included but based on different formalisms, a comparison of the resonance parameters extracted should
provide this information.

\begin{figure}
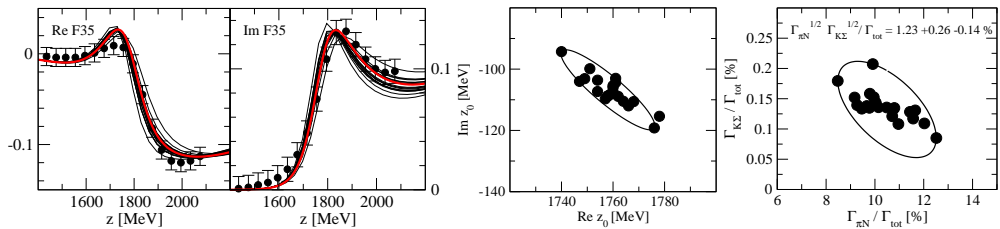

\begin{center}
\includegraphics[height=0.218\textwidth]{f35_errors.eps} \hspace*{-0.1cm}
\includegraphics[height=0.218\textwidth]{f35_pole_scatter_for_paper.eps} \hspace*{-0.1cm}
\includegraphics[height=0.218\textwidth]{f35_brara_scatter_for_paper.eps} 
\end{center}
\caption{Left: The F35 partial wave in $\pi N\to\pi N$. Data points: error given to the energy dependent
SAID solution~\cite{Arndt:2006bf}, as used for the numerical fit. Thick (red) line: Minimal $\chi^2$
solution. Thin (black) lines: Representative solutions in the $\chi^2+1$ criterion (determination of the
non-linear error). Center and right: The $\Delta(1905)F_{35}$ pole positions and branching ratios from those
solutions. The ellipses are introduced to guide the eye.}
\label{fig:errors}
\end{figure}

As mentioned in Sec.~\ref{sec:fit}, the error bars in the $\chi^2$ minimization have been taken  from
experiment for the reaction $\pi^+p\to K^+\Sigma^+$,  but assigned by hand for the partial waves of elastic
$\pi N$ scattering, since no uncertainties are provided for the energy dependent partial wave amplitudes
provided by the GWU/SAID analysis~\cite{Arndt:2006bf} and the uncertainties provided for the corresponding
energy independent analysis do not have direct statistical meaning. The uncertainties for the $\pi N$
partial waves are chosen such that the contributions from both reactions to the total $\chi^2$ are
approximately equal. This makes a rigorous error analysis of the present results impossible, which would
require a fit directly to the elastic $\pi N$ data. Nevertheless, assuming these assigned errors are
realistic, we outline in this section how to obtain in principle the uncertainties on the parameters and
derived quantities, like pole positions and residues. The error (0.01) and energy spacing (40 MeV) used to
include the $\pi N$ partial waves in the $\chi^2$ minimization are shown in Fig.~\ref{fig:errors} for the
example of the F35 partial wave.  
 
As mentioned in Sec.~\ref{sec:fit}, 40 parameters tied to the resonances have been varied to minimize the
$\chi^2$. There are also other parameters tied to the non-resonant part, given by the form factors shown in
Table~\ref{tab:cutoff_ky}. The latter have been roughly adjusted by hand before carrying out the numerical
fit of the resonance parameters, and we do not consider them as free parameters for the error analysis
carried out in the following. In the space of 40 parameters the error of parameter $p_i$ is determined by
the range of $p_i$ for which the best $\chi^2_{\rm min}$ rises by less than $\Delta\chi^2=1$, optimizing at
the same time all other 39 parameters. In the limit $\Delta\chi^2\to 0$, this non-linear parameter error
approaches the usual parabolic error that can be obtained, e.g., from the Hesse matrix. 

Here, we consider only the example of the F35 partial wave. We restrict the determination of the error to
the 5-parameter subspace tied to the F35 resonance, e.g., its bare mass [cf. Eq.~(\ref{2res})] and four
couplings to the channels $\pi N$, $\rho N$, $\pi\Delta$, and $K\Sigma$ [cf. Eq.~(\ref{dressed}),
\ref{sec:bare_res}]. Furthermore, we determine the parameter errors within this subspace. This means that in
the optimization involved in the determination of the parameter error (see above), only the 4 parameters of
the subspace are varied while leaving the other 35 at the optimum. We have checked that this restriction has
surprisingly little influence on the parameter error because partial waves are explicitly included in the
fit instead of $\pi N$ observables; for example, varying a resonance parameter of the P33 partial wave
influences a parameter error of the F35 partial wave only indirectly through the inclusion of the $K\Sigma$
data in the total $\chi^2$. 

The non-linear parameter errors are shown in Table~\ref{tab:parmerrors}.
\begin{table*}
 \caption{Error estimates of bare mass $m_b$ and bare couplings $f$ for the $\Delta (1905)F_{35}$ resonance.
          For other bare parameters see Table~\ref{bare_cou}.}
 \renewcommand{\arraystretch}{1.5}
 \begin{center}
 \begin{tabular}[t]{ccccc}
 \hline 
 \hline
 $m_{b}$ [MeV]   	& $\pi N$	 		& $\rho N$	 		& $\pi\Delta$   		& $\Sigma K$	 \\ 
 $2258^{+44}_{-43}$ 	& $0.0500^{+0.0011}_{-0.0012}$	& $-1.62^{+1.29}_{-1.61}$	& $-1.15^{+0.030}_{-0.022}$	& $0.120^{+0.0065}_{-0.0059}$	 \\
 \hline 
 \hline
 \end{tabular}
 \end{center}
 \label{tab:parmerrors}
 \end{table*}
The errors are small for the bare couplings to the $\pi N$ and $K\Sigma$ channels, for both of which the
data constrain the values. The errors are larger for the bare coupling to the $\rho N$ state that is less
constrained by data. Indeed, there is a strong correlation between the bare $\rho N$ coupling and the bare
mass, which therefore also has quite a large parameter error. Although no data are included for the
$\pi\Delta$ channel, the corresponding bare coupling has small errors. This is because the $\pi\Delta$
channel provides most of the $\pi\pi N$ phase space that is responsible for the inelastic resonance width,
which is well constrained by the elastic $\pi N$ amplitude as shown in Fig.~\ref{fig:errors} to the left.

From the non-linear parameter errors, one can determine the uncertainties of derived quantities such as pole
positions, residues, branching ratios, or the amplitude itself. To scan the parameter space within the
errors, we have taken four sample points of a given parameter within its error (always optimizing all other
parameters). As there are five parameters in the considered subspace, 20 solutions are obtained from which
the F35 amplitude and the $\Delta(1905)F_{35}$ pole position and residues have been calculated [see
Fig.~\ref{fig:errors} and Table~\ref{tab:reserror}]. The error on these quantities is then given by the
maximal range reached by these solutions.
\begin{table*}
 \caption{Error estimates of pole position and residues for the $\Delta (1905)F_{35}$ resonance.}
 \renewcommand{\arraystretch}{1.5}
 \begin{center}
 \begin{tabular}[t]{cc|ccc}
 \hline 
 \hline
 			&			&			& $\pi N\to\pi N$	& $\pi N\to K\Sigma$ 	\\
 Re $z_0$ [MeV]   	& $1764^{+18}_{-20}$	& $|r|$ [MeV]	 	& $11^{+1.7}_{-1.4}$	& $1.4^{+0.24}_{-0.21}$	\\
 Im $z_0$ [MeV] 	& $-109^{+13}_{-12}$	& $\theta$ [$^0$]	& $-45^{+3.8}_{-11}$	& $-313^{+4.2}_{-10}$	\\
 \hline 
 \hline
 \end{tabular}
 \end{center}
 \label{tab:reserror}
 \end{table*}
In Fig.~\ref{fig:errors}, we show also that there are correlations between real and imaginary part of the
pole position and also between the branching ratios. As mentioned before, the transition branching ratio is
indeed better determined than the individual branching ratio into $K\Sigma$; we obtain $(\Gamma_{\pi
N}^{1/2}\,\Gamma_{K\Sigma}^{1/2})/\Gamma_{\rm tot}=1.23^{+0.26}_{-0.14}\,\%$.

Note that the F35 amplitudes allowed by the discussed $\chi^2+1$ criterion, shown in Fig.~\ref{fig:errors}
to the left,  lead to a much larger rise $\Delta\chi^2_{\pi N, {\rm F35}}\gg 1$ in the $\chi^2_{\pi N, {\rm
F35}}$ of the $\pi N$ data alone, as an inspection by eye shows. Still, these solutions fulfill the
$\chi^2+1$ criterion due to the contribution from the $K\Sigma$ data to the total $\chi^2$. If one
determines, e.g., the uncertainty of the pole position or branching ratio from $\pi N$ data alone, one would
obtain, of course, much smaller errors on these quantities. Thus, the uncertainties on pole position and
branching ratio shown in Fig.~\ref{fig:errors} and Table~\ref{tab:reserror} should be understood as upper
limits.  

In summary, we have outlined how to determine the statistical errors of the present results, for the example
of the F35 resonance. A rigorous statistical analysis, as outlined above and carried out in
Ref.~\cite{Borasoy:2006sr} for $K^-p$ scattering, requires a direct fit to $\pi N$ observables and the full
inclusion of $K\Lambda$, $K\Sigma (I=1/2)$ and $\eta N$ data and will be carried out in the future, but the
present discussion serves to illustrate the error one expects from such an analysis.


\section{Summary}

A first combined analysis of the reactions $\pi N\to\pi N$ and $\pi^+ p\to K^+\Sigma^+$ within the
unitary dynamical coupled-channels framework has been presented. For the  $\pi^+ p\to K^+\Sigma^+$
reaction, the world data set from threshold to $z=2.3$ GeV has been considered. 

Dynamical coupled-channels models are particularly suited for combined data analyses: the SU(3) fla\-vor
symmetry for the exchange processes allows to relate different final states. The $t$- and $u$-channel
diagrams connect also different partial waves and the respective backgrounds. 

As a result, for both $\pi N$ and $K\Sigma$, a realistic and structured background can be provided,
depending only on a few free constants and form factors whose values are all in a natural range. 
Consequently, only a minimal set of bare $s$-channel resonances
is needed to obtain a good fit to the combined data sets.
This may also be tied to the fact that in this field-theoretical, Lagrangian based approach, the dispersive
parts from intermediate states are fully included and thus, analyticity is ensured.

Apart from the well-established 4-star resonances, a wide $\Delta(1600)P_{33}$ state has been found,
dynamically generated from the unitarization of the $t$- and $u$-channel exchanges. Furthermore, there is a
clear need for the three-star $\Delta(1920)P_{33}$ resonance. This state is found to couple only weakly to
$\pi N$ but stronger to $K\Sigma$. Thus, in the present combined analysis of elastic $\pi N$ scattering and
$K^+\Sigma^+$ production,  evidence for a ``missing resonance state''~\cite{Koniuk:1979vw} could be
accumulated which indeed has no clear signal in elastic $\pi N$ scattering alone.

\section*{Acknowledgement}

The work of M.D. is supported by DFG (Deutsche Forschungsgemeinschaft, GZ: DO 1302/1-2). This work is
supported in part by the Helmholtz Association through funds provided to the virtual institute ``Spin and
Strong QCD'' (VH-VI-231), by the  EU-Research Infrastructure Integrating Activity ``Study of Strongly
Interacting Matter" (HadronPhysics2, grant n. 227431) under the Seventh Framework Program of EU and by the
DFG (TR 16). F.H. is grateful to the COSY FFE grant No. 41788390 (COSY-058).


\appendix

\section{Exchange potentials with KY states}
\label{sec:su3}
In \ref{sec:explicit_exchanges} we list the explicit expressions for the $t$- and $u$-channel exchange
diagrams that involve the $K\Lambda$ and $K\Sigma$ channels. For the other exchange processes contained in
the model, see Ref.~\cite{Gasparyan:2003fp}. The new coupling constants for the exchange processes with $KY$
participation are related to the cases without $KY$ through SU(3) symmetry. The corresponding expressions
can be found in \ref{sec:su3_couplings}.

\begin{figure}
\begin{center}
\includegraphics[height=0.43\textwidth]{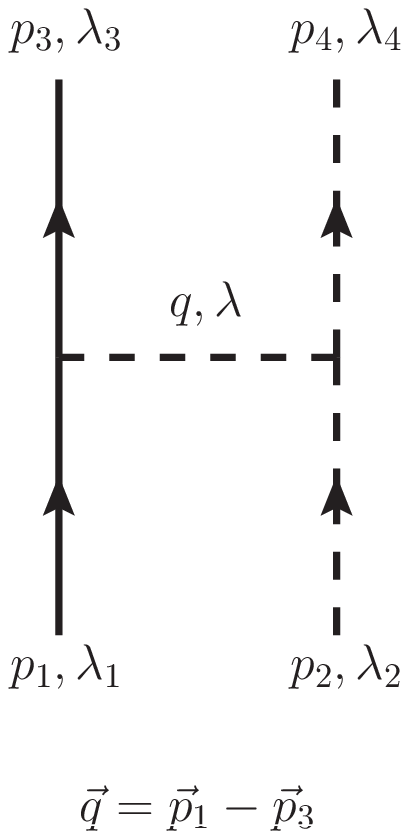} \hspace*{0.5cm}
\includegraphics[height=0.43\textwidth]{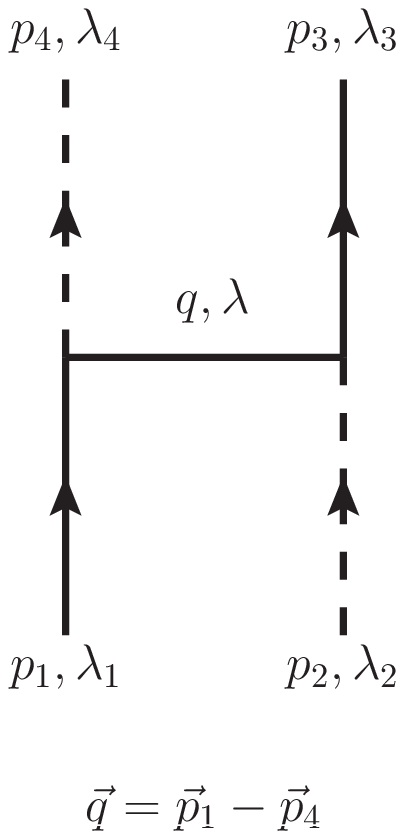}
\end{center}
\caption{$t$- and $u$-channel exchange processes.}
\label{fig:exchanges}
\end{figure}

The  kinematical quantities are specified in Fig.~\ref{fig:exchanges}. The index $1$ and $3$ ($2$ and $4$)
denote the incoming and outgoing baryon (meson).  The on-shell energies are
\begin{eqnarray}
 E_{i}=\sqrt{\vec{p}_{i}^{\,2}+m_{B,\,i}^{2}},\quad \omega_{i}=\sqrt{\vec{p}_{i}^{\,2}+m_{i}^{2}}
\end{eqnarray}
for the baryon and the meson, respectively. In the TOPT framework used in this study, the zeroth component
of the initial and final momenta are set to their on-mass-shell values: $p^{0}_{i}=E_{i}$ or
$p^{0}_{i}=\omega_{i}$.

$\vec{q}$ is the three-momentum of the intermediate particle. $q$ with  $q^{0}=E_{q}$ (baryon exchange) or
$q^{0}=\omega_{q}$ (meson exchange) means the 4-momentum in the first time ordering whereas $\tilde{q}$
indicates the second time ordering with $\tilde{q}^{0}=-E_{q}$ (baryon exchange) or
$\tilde{q}^{0}=-\omega_{q}$ (meson exchange).  Furthermore, in the potentials quoted in
\ref{sec:explicit_exchanges}, $P^{\mu\nu}$ is the  Rarita-Schwinger propagator of spin $3/2$ particles given
in Ref.~\cite{Krehl:1999km}, and $\hat p\equiv \gamma^\mu\,p_\mu$.

If the intermediate particle is a 
$\Sigma^*$ baryon, we use \cite{Schutz:1994wp} $q_{\Sigma*}^{0}=\epsilon_{1}-\epsilon_{4}$ with
\begin{eqnarray}
 \epsilon_{1}=\frac{z^{2}+m_{1}^{2}-m_{2}^{2}}{2\,z}\;
  ,\;\;\epsilon_{4}=\frac{z^{2}-m_{3}^{2}+m_{4}^{2}}{2\,z}
\end{eqnarray}
Each exchange diagram includes a kinematical normalization factor
\begin{eqnarray}
 \kappa=\frac{1}{(2\pi)^{3}}\frac{1}{2\sqrt{\omega_{2}\omega_{4}}}.
\end{eqnarray}
The isospin factors $IF$ for the exchange processes without the participation of the $KY$ channels can be
found in Refs.~\cite{Krehl:1999km,Gasparyan:2003fp}.  For the extension to the $KY$ channels, carried out in
the present work, the isospin factors can be found in Table~\ref{tab.: isospin2}. Also, every exchange
process quoted in \ref{sec:explicit_exchanges} is multiplied with form factors corresponding to the two
vertices, quoted in \ref{sec:su3_couplings}.

\begin{table}
\caption{Isospin factors for exchange diagrams with participation of the $KY$ channels. See
Refs.~\cite{Krehl:1999km,Gasparyan:2003fp}  for the corresponding values for the other diagrams.}
\begin{center}
\renewcommand{\arraystretch}{1.30}
\begin{tabular}[t]{lccc|lccc}
\hline\hline
  Transition & Exchange & IF$(\frac{1}{2})$ & IF$(\frac{3}{2})$ &  Transition & Exchange & IF$(\frac{1}{2})$ & IF$(\frac{3}{2})$ \\ \hline
$\pi N\rightarrow \Lambda K$&$\kst$ ex. &$\sqrt{3}$&$0$&$\Lambda K\rightarrow \Lambda K$&$ \sigma$ ex. &$1$&0\\
			&$\Sigma$ ex. &$\sqrt{3}$&$0$&&$ \omega$ ex. &$1$&0\\
			&$\Sigma^{*}$ ex. &$\sqrt{3}$&$0$&&$ \phi$ ex. &$1$&0\\ 
$\pi N\rightarrow \Sigma K$&$\kst$ ex. &$1$&$2$&$\Lambda K\rightarrow\Sigma K$&$\kst$ ex.&$-\sqrt{3}$&$0$\\
			&$\Sigma$ ex. &$2$&$1$&$\Sigma K\rightarrow \Sigma K$&$\sigma$ ex. &$1$&$1$\\
			&$\Lambda$ ex. &$-1$&$1$&&$\omega$ ex. &$1$&$1$\\
			&$\Sigma^{*}$ ex. &$2$&$1$&&$\phi$ ex. &$1$&$1$\\ 
&&&&&$\rho$ ex. &$2$&$-1$\\
 \hline \hline
\end{tabular}
\label{tab.: isospin2}
\end{center}
\end{table}


\subsection{Amplitudes for the exchange diagrams with $KY$}
\label{sec:explicit_exchanges}

The following expressions give the amplitudes shown in Fig.~\ref{fig:kydiagrams} in the helicity base, 
\be
\langle \lambda'|V(\vec{k},\vec{k}',z)|\lambda\rangle
\ee
with the notation for the momenta as in Fig.~\ref{fig:exchanges} and the dependence of the amplitude is on
the c.m.  (off-shell)  three-momentum of the incoming (outgoing) meson-baryon system, $\vec k\,(\vec k')$,
and the total scattering energy $z$. To solve the scattering equation in the $JLS$-basis, these
expressions still have to be partial wave projected~\cite{Schutz:1998jx,Krehl:1999km,Gasparyan:2003fp}.

\newpage

\vspace{0.3cm}

\underline{$\pi N\rightarrow K\Lambda$}
\begin{itemize}
 \item $\kst$ $t$-exchange
\end{itemize}
 
\begin{eqnarray}
&&\kappa g_{K\pi\kst}\, \bar{u}(\vec{p}_{3},\lambda_{3})
\Bigl([g_{\Lambda
N\kst}\gamma^{\mu}-i\frac{f_{\Lambda N
K^{*}}}{2m_{N}}\sigma^{\mu\nu}q_{\nu}]\,\frac{1}{z-\omega_{q}-E_{3}-\omega_{2}} \nonumber\\ 
&+& [g_{\Lambda
N\kst}\gamma^{\mu}-i\frac{f_{\Lambda N K^{*}}}{2m_{N}}\sigma^{\mu\nu}\tilde{q}_{\nu}]
\,\frac{1}{z-\omega_{q}-E_{1}-\omega_{4}} \Bigr)  \, u(\vec{p}_{1},\lambda_{1})
\frac{(p_{2}+p_{4})_{\mu}}{2\omega_{q}}\, IF \nonumber
\end{eqnarray}

\begin{itemize}
\item $\Sigma$ $u$-exchange
\end{itemize}
\begin{eqnarray}
\kappa \frac{g_{N\Sigma K}g_{\pi\Sigma\Lambda}}{m_{\pi}^{2}}\, \bar{u}(\vec{p}_{3},\lambda_{3})\gaf
\frac{\hat{p}_{2}}{2E_{q}} 
\left(\frac{\hat{q}+m_{\Sigma}}{z-E_{q}-\omega_{2}-\omega_{4}}
+\frac{\hat{\tilde{q}}+m_{\Sigma}}{z-E_{q}-E_{1}-E_{3}}\right)
\gaf \hat{p}_{4}
u(\vec{p}_{1},\lambda_{1})\, IF\nonumber
\end{eqnarray}

\begin{itemize}
\item $\Sigma^{*}$ $u$-exchange
\end{itemize}
\begin{eqnarray}
\kappa\frac{g_{N\Sigma^{*}
K}g_{\pi\Sigma^{*}\Lambda}}{m_{\pi}^{2}}\bar{u}(\vec{p}_{3},\lambda_{3})\,\frac{p_{2_{\mu}}
P^{\mu\nu}(q_{\Sigma*})
}{2E_{q}}\left(\frac{1}{z-E_{q}-\omega_{2}-\omega_{4}}+\frac{1}{z-E_{q}-E_{1}-E_{3}}
\right)
 p_{4_{\nu}}\,u(\vec{p}_{1},\lambda_{1})\, IF\nonumber
\end{eqnarray}

\underline{$\pi N\rightarrow K\Sigma$}

\begin{itemize}
\item
$\kst$ $t$-exchange
\end{itemize}

\begin{eqnarray}
&&\kappa g_{K\pi\kst}\, \bar{u}(\vec{p}_{3},\lambda_{3})
\Bigl([g_{\Sigma
N\kst}\gamma^{\mu}-i\frac{f_{\Sigma N
K^{*}}}{2m_{N}}\sigma^{\mu\nu}q_{\nu}]\frac{1}{z-\omega_{q}-E_{3}-\omega_{2}} 
\nonumber\\ &+& [g_{\Sigma
N\kst}\gamma^{\mu}-i\frac{f_{\Sigma N K^{*}}}{2m_{N}}\sigma^{\mu\nu}\tilde{q}_{\nu}]
\frac{1}{z-\omega_{q}-E_{1}-\omega_{4}} \Bigr) 
\, u(\vec{p}_{1},\lambda_{1})
\frac{(p_{2}+p_{4})_{\mu}}{2\omega_{q}}\, IF \nonumber
\end{eqnarray}

\begin{itemize}
\item $\Sigma$ $u$-exchange
\end{itemize}
\begin{eqnarray}
\kappa \frac{g_{N\Sigma K}g_{\pi\Sigma\Sigma}}{m_{\pi}^{2}}\, \bar{u}(\vec{p}_{3},\lambda_{3})\gaf
\frac{\hat{p}_{2}}{2E_{q}} 
\left(\frac{\hat{q}+m_{\Sigma}}{z-E_{q}-\omega_{2}-\omega_{4}}
+\frac{\hat{\tilde{q}}+m_{\Sigma}}{z-E_{q}-E_{1}-E_{3}}\right)\
 \gaf \hat{p}_{4}
u(\vec{p}_{1},\lambda_{1})\, IF\nonumber
\end{eqnarray}

\begin{itemize}
\item $\Lambda$ $u$-exchange
\end{itemize}
\begin{eqnarray}
\kappa \frac{g_{N\Lambda K}g_{\pi\Lambda\Sigma}}{m_{\pi}^{2}}\, \bar{u}(\vec{p}_{3},\lambda_{3})\gaf
\frac{\hat{p}_{2}}{2E_{q}} 
\left(\frac{\hat{q}+m_{\Lambda}}{z-E_{q}-\omega_{2}-\omega_{4}}
+\frac{\hat{\tilde{q}}+m_{\Lambda}}{z-E_{q}-E_{1}-E_{3}}\right)
\gaf \hat{p}_{4}
u(\vec{p}_{1},\lambda_{1})\, IF\nonumber
\end{eqnarray}

\begin{itemize}
\item $\Sigma^{*}$ $u$-exchange
\end{itemize}
\begin{eqnarray}
\kappa\frac{g_{N\Sigma^{*}
K}g_{\pi\Sigma^{*}\Sigma}}{m_{\pi}^{2}}\bar{u}(\vec{p}_{3},\lambda_{3})\,\frac{p_{2_{\mu}}
P^{\mu\nu}(q_{\Sigma*})
}{2E_{q}}\left(\frac{1}{z-E_{q}-\omega_{2}-\omega_{4}}+\frac{1}{z-E_{q}-E_{1}-E_{3}}
\right)
 p_{4_{\nu}}\,u(\vec{p}_{1},\lambda_{1})\, IF\nonumber
\end{eqnarray}

\newpage

\underline{$K\Lambda\rightarrow K\Lambda$}
\begin{itemize}
 \item $\sigma$ $t$-exchange
\end{itemize}
\begin{eqnarray}
\kappa\frac{g_{\sigma \Lambda\Lambda}g_{\sigma K
K}}{2m_{\pi}}\frac{(-2p_{2_{\mu}}p^{\mu}_{4})}{2\omega_{q}}
\left(\frac{1}{z-E_{q}-E_{3}-\omega_{2}}+\frac{1}{z-E_{q}-E_{1}-\omega_{4}} \right)
\bar{u}(\vec{p}_{3},\lambda_{3})\,u(\vec{p}_{1},\lambda_{1})\, IF\nonumber
\end{eqnarray}

\begin{itemize}
\item $\omega$ $t$-exchange
\end{itemize}
\begin{eqnarray}
&&\kappa g_{\omega KK}\, \bar{u}(\vec{p}_{3},\lambda_{3})
\Bigl([g_{\Lambda\Lambda\omega}\gamma^{\mu}- i\frac{f_{\Lambda\Lambda\omega}}{2m_{N}}
\sigma^{\mu\nu}q_{\nu}]\frac{1}{z-\omega_{q}-E_{3}-\omega_{2}} 
\nonumber\\ &+&
[g_{\Lambda\Lambda\omega}\gamma^{\mu}-i \frac{f_{\Lambda\Lambda\omega}}{2m_{N}}
\sigma^{\mu\nu}\tilde{q}_{\nu}] \frac{1}{ z-\omega_{q}-E_{1}-\omega_{4}} \Bigr)
u(\vec{p}_{1},\lambda_{1}) \frac{ (p_{2}+p_{4})_{\mu}}{2\omega_{q}}\, IF \nonumber
\end{eqnarray}

\begin{itemize}
\item $\phi$ $t$-exchange
\end{itemize}
\begin{eqnarray}
&&\kappa g_{\phi KK}\, \bar{u}(\vec{p}_{3},\lambda_{3})
\Bigl([g_{\Lambda
\Lambda\phi}\gamma^{\mu}-i \frac{f_{\Lambda\Lambda\phi}}{
2m_{N}}\sigma^{\mu\nu}q_{\nu}]\frac{1}{z-\omega_{q}-E_{3}-\omega_{2}} 
\nonumber\\ &+&
[g_{\Lambda\Lambda\phi} \gamma^{\mu}-i\frac{f_{\Lambda\Lambda\phi}}{2m_{N}}\sigma^{\mu\nu}\tilde{q}_{\nu}]
\frac{1}{z-\omega_{q}-E_{1}- \omega_{4}} \Bigr) \, u(\vec{p}_{1},\lambda_{1})
\frac{(p_{2}+p_{4})_{\mu}}{ 2\omega_{q}}\, IF \nonumber
\end{eqnarray}

\underline{$K\Lambda\rightarrow K\Sigma$}
\begin{itemize}
 \item $\rho$ $t$-exchange
\end{itemize}
\begin{eqnarray}
&&\kappa g_{K\rho K}\, \bar{u}(\vec{p}_{3},\lambda_{3})
\Bigl([g_{\Lambda\rho
\Sigma}\gamma^{\mu}-i\frac{f_{\Lambda\rho\Sigma}}{
2m_{N}}\sigma^{\mu\nu}q_{\nu}]\frac{1}{z-\omega_{q}-E_{3}-\omega_{2}} 
\nonumber\\ &+&
[g_{\Lambda\rho\Sigma} \gamma^{\mu}-i\frac{f_{\Lambda\rho\Sigma}}{2m_{N}}\sigma^{\mu\nu}\tilde{q}_{\nu}]
\frac{1}{z-\omega_{q}-E_{1}-\omega_{4}} \Bigr) 
\, u(\vec{p}_{1},\lambda_{1})
\frac{(p_{2}+p_{4})_{\mu}}{2\omega_{q}}\, IF \nonumber
\end{eqnarray}

\underline{$K\Sigma\rightarrow K\Sigma$}

\begin{itemize}
 \item $\sigma$ $t$-exchange
\end{itemize}
\begin{eqnarray}
\kappa\frac{g_{K\sigma
K}g_{\Sigma\sigma\Sigma}}{2m_{\pi}}\frac{(-2p_{2_{\mu}}p^{\mu}_{4})}{2\omega_{q}}
\left(\frac{1}{z-E_{q}-E_{3}-\omega_{2}}+\frac{1}{z-E_{q}-E_{1}-\omega_{4}} \right)
\bar{u}(\vec{p}_{3},\lambda_{3})\,u(\vec{p}_{1},\lambda_{1})\, IF\nonumber
\end{eqnarray}

\begin{itemize}
\item $\omega$ $t$-exchange
\end{itemize}
\begin{eqnarray}
&&\kappa g_{\omega KK}\, \bar{u}(\vec{p}_{3},\lambda_{3})
\Bigl([g_{\Sigma
\Sigma\omega}\gamma^{\mu}-i\frac{f_{
\Sigma\Sigma\omega}}{2m_{N}}\sigma^{\mu\nu}q_{\nu}]\frac{1}{z-\omega_{q}-E_{3}-\omega_{2}} 
\nonumber\\ &+&
[g_{\Sigma\Sigma\omega}\gamma^{\mu}-i\frac{f_{\Sigma\Sigma\omega}}{2m_{N}}\sigma^{\mu\nu}\tilde{q}_{\nu}]
\frac{1}{z-\omega_{q}-E_{1}-\omega_{4}} \Bigr) 
\, u(\vec{p}_{1},\lambda_{1})
\frac{(p_{2}+p_{4})_{\mu}}{2\omega_{q}}\, IF \nonumber
\end{eqnarray}

\newpage

\begin{itemize}
\item $\phi$ $t$-exchange
\end{itemize}
\begin{eqnarray}
&&\kappa g_{\phi KK}\, \bar{u}(\vec{p}_{3},\lambda_{3})
\Bigl([g_{
\Sigma\Sigma\phi}\gamma^{
\mu}-i\frac{f_{\Sigma\Sigma\phi}}{2m_{N}}\sigma^{\mu\nu}q_{\nu}]\frac{1}{z-\omega_{q}-E_{3}-\omega_{2}}
\nonumber\\ &+&
[g_{\Sigma\Sigma\phi}\gamma^{\mu}-i\frac{f_{\Sigma\Sigma\phi}}{2m_{N}}\sigma^{\mu\nu}\tilde{q}_{\nu}]
\frac{1}{z-\omega_{q}-E_{1}-\omega_{4}} \Bigr) \, u(\vec{p}_{1},\lambda_{1})
\frac{(p_{2}+p_{4})_{\mu}}{2\omega_{q}}\, IF \nonumber
\end{eqnarray}

\begin{itemize}
\item $\rho$ $t$-exchange
\end{itemize}
\begin{eqnarray}
&&\kappa g_{\rho KK}\, \bar{u}(\vec{p}_{3},\lambda_{3})
\Bigl([g_{\Sigma
\Sigma\rho}\gamma^{\mu}-i\frac{ f_{\Sigma\Sigma\rho}}{2m_{N}}
\sigma^{\mu\nu}q_{\nu}]\frac{1}{z-\omega_{q}-E_{3}-\omega_{2}} 
\nonumber\\ &+&
[g_{\Sigma\Sigma\rho}\gamma^{\mu}-i\frac{f_{\Sigma\Sigma\rho}}{2m_{N}}\sigma^{\mu\nu}\tilde{q}_{\nu}]
\frac{1}{z-\omega_{q}-E_{1}-\omega_{4}} \Bigr) \, u(\vec{p}_{1},\lambda_{1})
\frac{(p_{2}+p_{4})_{\mu}}{2\omega_{q}}\, IF \nonumber
\end{eqnarray}


\subsection{Coupling constants, form factors ($t$-, $u$-exchanges)}
\label{sec:su3_couplings}

First, the coupling constants needed for the exchange diagrams (Fig.~\ref{fig:kydiagrams}, expressions in 
\ref{sec:explicit_exchanges}) are quoted, followed by the form factors and their values. 

The coupling constants are related to their counterparts without strange particles by SU(3) flavor
symmetry~\cite{de Swart:1963gc} as outlined in Sec.~\ref{sec:newpotentials}. 
The coupling $8_B\otimes 8_B\otimes 8_M$ of baryon octets
$8_B$ and pseudoscalar meson octet $8_M$ depends on two parameters $g_1,
\,g_2$ corresponding to the symmetric and antisymmetric octet. They can be related to the coupling constant
$g$ and the mixing parameter $\alpha$ in the notation of Ref.~\cite{de Swart:1963gc}, which is also used
here, 	
\begin{eqnarray}
 g=\frac{\sqrt{30}}{40}\,g_{1}+\frac{\sqrt{6}}{24}\,g_{2},\quad \alpha = 
\frac{\sqrt{6}}{24}\,\frac{g_{2}}{g}. 
\end{eqnarray}
$g_{1}$ and $g_{2}$ can be also expressed in terms of the standard $D$ and $F$ couplings \cite{pdg}:
\begin{eqnarray}
 D=\frac{\sqrt{30}}{40}g_{1}\,  , \;\;\; F= \frac{\sqrt{6}}{24}\,g_{2}
\end{eqnarray}
with $\alpha=F/(D+F)$ (definition of $\alpha$ of Ref.~\cite{de Swart:1963gc}). The couplings of the physical
$\omega$ and $\phi$ are obtained assuming ideal mixing of the SU(3) states $\omega_1$, $\omega_8$, i.e., the
$\phi$ meson does not couple to the nucleon (see, e.g., Ref.~\cite{pdg}). 

Within SU(3), the mixing angle $\alpha$ is a free parameter but can be determined from SU(6). 
We use \cite{Janssen:1996kx} (The index $P$ denotes a pseudoscalar meson and $V$ means a vector meson) 
\begin{eqnarray}
 \alpha_{BBP}= 0.4\;, \quad
\alpha_{BBV}= 1.15\;, \quad
\alpha_{PPV}= 1 \;.
\end{eqnarray}
For the coupling of vector mesons to octet baryons we use the following Lagrangian:
\begin{eqnarray}
 \mathcal{L}_{int}=-g_{NN\rho}\bar\Psi[\gam-\frac{\kappa_{\rho}}{2m_{N}}\sigma^{\mu\nu}\partial_{\nu}]
\vtau\, \vec{\rho}_{\mu}\, \Psi 
\end{eqnarray}
which consists of a vector part with $\gam$ and a tensor part with $\sigma^{\mu\nu}$. The three fields
$\bar\Psi$, $\Psi$ and $\vec{\rho}$ are connected through a vector coupling $g_{NN\rho}$ and a tensor
coupling $f_{BBV}=g_{NN\rho}\kappa_{\rho}$~\cite{Holzenkamp:1989tq, Reuber:1993ip}.  The SU(3) couplings
quoted below are therefore divided into a vector part and a tensor part. We use
$\kappa_{\rho}=6.1$~\cite{Janssen:1996kx} and $f_{NN\omega}=0$.

The coupling of the $\Lambda$ and the $\Sigma$ to the $\sigma$ meson was not fixed by SU(3) symmetry since
the $\sigma$ is not a stable particle but regarded as an effective meson-meson state~\cite{Schutz:1998jx,
Krehl:1999km, Gasparyan:2003fp}, i.e. as a correlated $\pi\pi$ exchange process. In the calculation for the
background diagrams we take the following values from \cite{MuellerGroeling:1990cw, Holzenkamp:1989tq},
obtained from the hyperon nucleon interaction:
\begin{eqnarray}
g_{\Lambda\Lambda\sigma}= 8.175\;, \quad
g_{KK\sigma}= 1.336.
\end{eqnarray}
The coupling $g_{\Sigma\Sigma\sigma}$ was set to a different value compared to \cite{
MuellerGroeling:1990cw,Holzenkamp:1989tq}. We choose $ g_{\Sigma\Sigma\sigma}= 29.657$. 
The SU(3) symmetric coupling constants are given by the following expressions:

\begin{itemize}
\item Couplings for octet baryon, octet baryon and pseudoscalar meson:
\begin{eqnarray}
 g_{NN\pi}&=& g_{BBP}\;, \nonumber\\
g_{\Sigma NK}&=& g_{BBP}(1-2\alpha_{BBP})\;,\nonumber\\
g_{\Lambda NK}&=& -\frac{1}{3}\sqrt{3}\, g_{BBP} (1+2\alpha_{BBP})\;,\nonumber\\
g_{\Lambda\Sigma \pi}&=&\frac{2}{\sqrt{3}}\,g_{BBP}\,(1-\alpha_{BBP})\;.
\label{2.21}
\end{eqnarray}

\item Vector coupling for octet baryon, octet baryon and vector meson:
\begin{eqnarray}
g_{NN\rho}&=&g_{BBV}\;,\nonumber\\
g_{NN\omega}&=& g_{BBV}(4\alpha_{BBV}-1)\;,\nonumber\\
g_{\Lambda N\kst}&=& -g_{BBV}\tfrac{1}{\sqrt{3}}(1+2\alpha_{BBV})\;,\nonumber\\
g_{\Sigma N\kst}&=&g_{BBV}(1-2\alpha_{BBV})\;,\nonumber\\ 
g_{\Lambda\Lambda\omega}&=& g_{BBV}\tfrac{2}{3}(5\alpha_{BBV}-2)\;,\nonumber\\
g_{\Sigma\Sigma\omega}&=&g_{BBV}2\alpha_{BBV}\;,\nonumber\\
g_{\Lambda\Lambda\phi}&=& -g_{BBV}\tfrac{\sqrt{2}}{3}(2\alpha_{BBV}+1)\;,\nonumber\\
g_{\Sigma\Sigma\phi}&=& -g_{BBV}\sqrt{2}(2\alpha_{BBV}-1)\;,\nonumber\\
g_{\Sigma\Sigma\rho}&=&g_{BBV}2\alpha_{BBV}\;,\nonumber\\
g_{\Lambda\Sigma\rho}&=&g_{BBV}\tfrac{2}{\sqrt{3}}(1-\alpha_{BBV})\;.
\label{2.22}
\end{eqnarray}

\item Tensor coupling for octet baryon, octet baryon and vector meson: 
\begin{eqnarray}
f_{NN\rho}&=& g_{NN\rho}\kappa_{\rho}\;,\nonumber\\
f_{\Lambda N\kst}&=& -f_{NN\omega}\tfrac{1}{2\sqrt{3}}-f_{NN\rho}\tfrac{\sqrt{3}}{2}\;,\nonumber\\
f_{\Sigma N\kst}&=& -f_{NN\omega}\tfrac{1}{2}+f_{NN\rho}\tfrac{1}{2}\;,\nonumber\\
f_{\Lambda\Lambda \omega}&=& f_{NN\omega}\tfrac{5}{6}-f_{NN\rho}\tfrac{1}{2}\;,\nonumber\\
f_{\Sigma\Sigma\omega}&=& f_{NN\omega}\tfrac{1}{2}+f_{NN\rho}\tfrac{1}{2}\;,\nonumber\\
f_{\Lambda\Lambda\phi}&=& -f_{NN\omega} \tfrac{1}{3\sqrt{2}}-f_{NN\rho}\tfrac{1}{\sqrt{2}}\;,\nonumber\\
f_{\Sigma\Sigma\phi}&=& -f_{NN\omega}\tfrac{1}{\sqrt{2}}+f_{NN\rho}\tfrac{1}{\sqrt{2}}\;,\nonumber\\
f_{\Sigma\Sigma\rho}&=& f_{NN\omega}\tfrac{1}{2}+f_{NN\rho}\tfrac{1}{2}\;,\nonumber\\
f_{\Lambda\Sigma\rho}&=&-f_{NN\omega}\tfrac{1}{2\sqrt{3}}+f_{NN\rho}\tfrac{\sqrt{3}}{2}\;.  \label{2.23}
\end{eqnarray}

\item Coupling for pseudoscalar meson, pseudoscalar meson and vector meson:
\begin{eqnarray}
g_{\pi\pi\rho}&=& 2\,g_{PPV}\;,\nonumber\\
g_{KK\rho}&=&g_{PPV}\;,\nonumber\\
g_{K\pi\kst}&=& -g_{PPV}\;,\nonumber\\
g_{KK\omega}&=& g_{PPV}\;,\nonumber\\
g_{KK\phi}&=&\sqrt{2}\, g_{PPV}\;. \label{2.24}
\end{eqnarray}

\item Coupling for decuplet baryon, octet baryon and pseudoscalar meson:
\begin{eqnarray}
g_{\Delta N\pi}&=& g_{DBP}\;,\nonumber\\
g_{\Sigma^{*} NK}&=& -g_{DBP}\tfrac{1}{\sqrt{6}}\;,\nonumber\\
g_{\Sigma^{*}\Sigma\pi}&=&g_{DBP}\tfrac{1}{\sqrt{6}}\;,\nonumber\\
g_{\Sigma^{*}\Lambda\pi}&=& g_{DBP}\tfrac{1}{\sqrt{2}}\;.
\end{eqnarray}

\end{itemize}

For the new diagrams of Fig.~\ref{fig:kydiagrams}, we use the same expressions for the form factors as in
Ref.~\cite{Gasparyan:2003fp},  
\begin{eqnarray} F(q)&=&
\left(\frac{\Lambda^{2}-m_{x}^{2}}{\Lambda^{2}+\vec{q}^{\,2}} \right)^{n} 
\end{eqnarray} 
where $m_{x}$ is the mass and $\vec{q}$ the momentum of the exchanged particle.  $n=1\,,\,2$ denotes a
monopole or dipole form factor. The dipole type applies to vertices with $\rho$, $\Delta$, $\kst$ or
$\Sigst$ as exchanged particle, otherwise a monopole form factor is used. For the form factors of all other
exchange diagrams of the model, see Ref.~\cite{Gasparyan:2003fp}. The numerical values for the new form
factors are given in Table~\ref{tab:cutoff_ky}.
\begin{table}
\caption{Form factors $\Lambda$ for the exchange diagrams with $KY$. The columns {\it Ex} specify the
exchanged particle. For the numerical values of the form factors of the other diagrams, see
Ref.~\cite{Gasparyan:2003fp}.}
\begin{center}
\renewcommand{\arraystretch}{1.50}
\begin{tabular}[t]{ccc|ccc}
\hline \hline 
Vertex 			& Ex 		& $\Lambda$ [MeV]	&Vertex 		& Ex 		& $\Lambda$ [MeV] \\ \hline 
$\pi\kst K $ 		& $\kst $	& $1700$		&$K\sigma K $ 		& $\sigma $	& $1400$ \\
$N\kst \Lambda $ 	& $\kst $   	& $1200$	       	&$K\omega K $	       	& $\omega $     & $1600$ \\
$N\kst\Sigma $ 		& $ \kst$  	& $1800$	       	&$K\phi K $	       	& $\phi $       & $1500$ \\
$\pi\Sigma\Lambda $ 	& $ \Sigma$  	& $1800$	       	&$\Lambda\sigma\Lambda$ & $\sigma $     & $1000$ \\
$\pi\Sigst\Lambda $ 	& $\Sigst $   	& $2000$	       	&$\Lambda\omega\Lambda$ & $\omega $     & $2000$ \\
$\pi\Sigma\Sigma $ 	& $\Sigma $    	& $1800$	       	&$\Lambda\phi\Lambda $  & $\phi $       & $1500$ \\
$\pi\Sigst\Sigma $ 	& $ \Sigst$   	& $2000$	       	&$\Lambda\rho\Sigma $   & $\rho $       & $1160$ \\
$N\Sigma K $ 		& $\Sigma $  	& $1800$	       	&$\Sigma\sigma\Sigma $  & $\sigma $     & $1000$ \\
$N\Sigst K $ 		& $\Sigst $   	& $2000$	       	&$\Sigma\omega\Sigma $  & $\omega $     & $2000$ \\
$\pi\Lambda \Sigma $ 	& $\Lambda $  	& $1800$	       	&$\Sigma\phi\Sigma $    & $\phi $       & $1600$ \\
$N\Lambda K $ 		& $\Lambda $ 	& $1800$	       	&$\Sigma\rho\Sigma $    & $\rho $       & $1350$ \\
\hline \hline
\end{tabular}
\label{tab:cutoff_ky}
\end{center}
\end{table} 
The numerical values for the other diagrams (without $KY$) have not been changed, see
Ref.~\cite{Gasparyan:2003fp} for the values.


\section{Bare resonance vertices}
\label{sec:bare_res}

The bare resonance vertices for $J\le 3/2$ are given by the effective Lagrangians listed in
Table~\ref{lbare}.  There, $\vec{\rho}_{\mu\nu}=\delmu\vec{\rho}_{\nu}-\delnu\vec{\rho}_{\mu}$ for the field
of the $\rho$ meson.  The vertices derived from these Lagrangians are partial wave projected to the
$JLS$-basis~\cite{Krehl:1999km,Gasparyan:2003fp}. 

The Lagrangians for the $K\Lambda$ couplings have the same structure as for the $\eta N$ couplings, the ones
for the $K\Sigma$ couplings have the same structure like for the $\pi N$ case except for the replacements
$\vec \tau\, \vec\pi\to\vec \tau \,\vec\Psi_{\Sigma}$  , $\vec{S}^{\dagger}\,\vec\pi\to\vec {S}^{\dagger}\,
\vec\Psi_{\Sigma}$, or $\vec{S}\,\vec\pi\to\vec {S}\, \vec\Psi_{\Sigma}$. Thus, they are not
quoted explicitly in Table~\ref{lbare}.

\begin{table}
\caption{Effective Lagrangians for the resonance vertices.}
\begin{center}
\renewcommand{\arraystretch}{1.40}
\begin{tabular}[t]{clcl}
\hline \hline
Vertex & $\mathcal{L}_{int}$ & Vertex & $\mathcal{L}_{int}$\\ \hline
$\Nst(S_{11})N\pi$& $ \frac{f}{\mpi}\bar{\Psi}_{\Nst}\gam\vtau\,\delpi\,\Psi \;+\;\text{h.c.}$ &
$\Dst(S_{31})N\pi $& $\frac{f}{\mpi}\bar{\Delta}^{*}\gam\vec{S}^{\dagger}\delpi\,\Psi\;+\;\text{h.c.}$ \\
$\Nst(S_{11})N\eta $& $\frac{f}{\mpi}\bar{\Psi}_{\Nst}\gam\partial_{\mu}\eta\,\Psi \;+\;\text{h.c.}$ &
$\Dst(S_{31})N\rho$ & $f \bar{\Delta}^{*}\vec{S}^{\dagger}\gaf\,\gam\,\vec{\rho}_{\mu}\Psi+\;\text{h.c.}$\\
$\Nst(S_{11})N\rho $& $f\,\bar{\Psi}_{\Nst}\gaf\gam\vtau\,\vec{\rho}_{\mu}\,\Psi \;+\;\text{h.c.}$ &
$\Dst(S_{31})\Delta\pi $& $\frac{f}{\mpi}\bar{\Delta}^{*}\gaf\vec{T}\delpi\,\Delta^{\mu}\;+\;\text{h.c.}$ \\
$\Nst(S_{11})\Delta\pi $& $\frac{f}{\mpi}\bar{\Psi}_{\Nst}\gaf\vec{S}\delpi\,\Delta^{\mu} \;+\;\text{h.c.}$ &
$\Dst(P_{31})N\pi $& $-\frac{f}{\mpi}\bar{\Delta}^{*}\gaf\gam\vec{S}^{\dagger}\delpi\,\Psi\;+\;\text{h.c.}$ \\
$\Nst(P_{11})N\pi $& $-\frac{f}{\mpi}\bar{\Psi}_{\Nst}\gaf\gam\vec{\tau}\,\delpi\,\Psi+\;\text{h.c.}$ &
$\Dst(P_{31})N\rho $ & $-f\, \bar{\Delta}^*\,\vec{S}^{\dagger}\, \gamma^{\mu}\,\vec{\rho}_{\mu}\Psi+\;\text{h.c.}$ \\
$\Nst(P_{11})N\eta $& $-\frac{f}{\mpi}\bar{\Psi}_{\Nst}\gaf\gam\,\partial_{\mu}\eta\,\Psi+\;\text{h.c.}$ &
$\Dst(P_{31})\Delta\pi $& $\frac{f}{\mpi}\bar{\Delta}^{*}\vec{T}\delpi\,\Delta^{\mu}\;+\;\text{h.c.}$ \\
$\Nst(P_{11})N\rho $ & $-f\, \bar{\Psi}_{\Nst}\, \gamma^{\mu}\,\vec{\tau}\,\vec{\rho}_{\mu}\Psi+\;\text{h.c.}$ &
$\Dst(P_{33})N\pi $ & $ \frac{f}{m_\pi}\bar{\Delta}^*_\mu \vec{S}^\dag  \partial^\mu\vec{\pi}\,\Psi +{\rm h.c.},$ \\
$\Nst(P_{11})\Delta\pi $& $\frac{f}{\mpi}\,\bar{\Psi}_{\Nst}\vec{S}\delpi\,\Delta^{\mu}\;+\;\text{h.c.}$ &
$\Dst(P_{33})N\rho $& $-i\frac{f}{m_{\rho}}\bar{\Delta}^{*}_{\mu}\gaf\gamma_{\nu}\vec{S}^{\dagger}\vec{\rho}^{\,\mu\nu}\,\Psi\;+\;\text{h.c.}$ \\
$\Nst(P_{13})N\pi $& $\frac{f}{\mpi}\bar{\Psi}_{\Nst}^{\mu}\vtau\,\delpi\,\Psi\;+\;\text{h.c.}$ &
$\Dst(P_{33})\Delta\pi $& $\frac{f}{\mpi}\bar{\Delta}^{*}_{\mu}\gaf\gan\,\vec{T}\delnu\vec{\pi}\,\Delta^{\mu}\;+\;\text{h.c.}$\\
$\Nst(P_{13})N\eta $& $\frac{f}{\mpi}\bar{\Psi}_{\Nst}^{\mu}\delmu\eta\,\Psi\;+\;\text{h.c.}$ &
$\Dst(D_{33})N\pi $& $\frac{f}{\mpi^{2}}\,\bar{\Delta}^{*}_{\mu}\,\gaf\,\gamma_{\nu}\vec{S}^{\dagger}\partial^{\nu}\partial^{\mu}\vec{\pi}\,\Psi+\;\text{h.c.}$ \\
$\Nst(P_{13})N\rho $&$-i\frac{f}{m_{\rho}}\bar{\Psi}_{\Nst}^{\mu}\gaf\gamma^{\nu}\vec{\tau}\,\vec{\rho}_{\,\mu\nu}\Psi\;+\;\text{h.c.}$ &
$\Dst(D_{33})\Delta\pi $& $i\,\frac{f}{\mpi}\bar{\Delta}^{*}_{\nu}\vec{T}\gam\delpi\,\Delta^{\nu}\;+\;\text{h.c.}$ \\
$\Nst(P_{13})\Delta\pi $& $\frac{f}{\mpi}\bar{\Psi}_{\Nst}^{\mu}\gaf\gan\vec{S}\delnu\vec{\pi}\,\Delta_{\mu}\;+\;\text{h.c.}$ &
$\Dst(D_{33})N\rho $& $\frac{f}{m_{\rho}}\bar{\Delta}^{*}_{\mu}\gamma_{\nu}\vec{S}^{\dagger}\vec{\rho}^{\,\mu\nu}\,\Psi\;+\;\text{h.c.}$\\ 
$\Nst(D_{13})N\pi $& $\frac{f}{\mpi^{2}}\bar{\Psi}\gaf\gan\vtau\,\delnu\delpi\,\Psi_{\Nst}^{\mu}\;+\;\text{h.c.}$ & \\
$\Nst(D_{13})N\eta $& $\frac{f}{\mpi^{2}}\bar{\Psi}\gaf\gan\,\delnu\delmu\eta\,\Psi_{\Nst}^{\mu}\;+\;\text{h.c.}$ & \\
$\Nst(D_{13})N\rho $& $\frac{f}{m_{\rho}}\bar{\Psi}^{\mu}_{\Nst}\gan\vtau\,\vec{\rho}_{\mu\nu}\Psi\;+\;\text{h.c.}$ & \\
$\Nst(D_{13})\Delta\pi $& $i\frac{f}{\mpi}\bar{\Psi}^{\nu}_{\Nst}\vec{S}\gam\delpi\,\Delta_{\nu}\;+\;\text{h.c.}$ & \\

\hline \hline
\end{tabular}
\end{center}
\label{lbare}
\end{table} 
Additionally, form factors are supplied for each vertex, given by
\begin{eqnarray}
F(k)&=& \left(\frac{\Lambda^{4}+m_{R}^{4}}{\Lambda^{4}+(E(k)+\omega(k))^{4}} \right)^{n}
\label{fofa}
\end{eqnarray}
where $m_{R}$ is the nominal mass of the resonance ($\sim {\rm Re}\, z_0$) and $E(k)$, $\omega(k)$ denote
the on-shell energies of the incoming or outgoing baryon and meson with c.m. (off-shell) momentum $k$. We
have $n=1$ in the case of $J\leq 3/2$ for all channels except for $\Delta\pi$ ($n=2$). For $J\geq 5/2$ we
have $n=2$ for all channels except for $\Delta\pi$ ($n=3$). For the cut-off parameter we choose $\Lambda=2$
GeV, except for the $\Delta(1232)P_{33}$ resonance, where we fine-tune $\Lambda=1.8,\,1.7$ GeV for the
vertices to the $\pi N$, $\pi\Delta$ states, respectively.

%
%

\begin{table*}
 \caption{Bare resonance parameters: masses $m_b$ and coupling constants $f$.}
 \renewcommand{\arraystretch}{1.3}
 \begin{center}
 \begin{tabular}[t]{cccccc}
 \hline 
 \hline
		      &	$m_{b}$ [MeV]	& $\pi N$	& $\rho N$ 	& $\pi\Delta$	& $\Sigma K$	\\ 
 \hline
 $\Delta (1232)P_{33}$&	$1535$		& $1.44$	& $5.88$	& $-0.551$	& $0.0316$	\\
 $\Delta (1620)S_{31}$&	$3669$		& $0.769$	& $1.107$	& $-6.05$	& $2.25$	\\
 $\Delta (1700)D_{33}$&	$3442$		& $0.100$	& $-6.47$	& $-0.845$    	& $0.170$	\\
 $\Delta (1905)F_{35}$&	$2258$		& $0.0500$	& $-1.62$	& $-1.15$	& $0.120$	\\
 $\Delta (1910)P_{31}$&	$3114$		& $0.367$	& $4.36$	& $-0.355$	& $0.231$	\\
 $\Delta (1920)P_{33}$&	$2508$		& $-0.123$	& $-2.96$	& $-0.530$	& $-1.86$	\\
 $\Delta (1930)D_{35}$&	$2332$		& $0.177$	& $-4.19$	& $-0.178$	& $4.12$	\\
 $\Delta (1950)F_{37}$&	$2597$		& $0.580$	& $12.3$	& $1.87$       	& $0.663$	\\
 \hline 
 \hline
 \end{tabular}
 \end{center}
 \label{bare_cou}
 \end{table*}

\begin{table}
\caption{Isospin factors $I_R$ for resonances vertices.}
\begin{center}
\renewcommand{\arraystretch}{1.4}
\begin{tabular}[t]{ccccccc}
\hline \hline
  		& $N\pi$ 	& $N\eta$ & $N\rho$ 	& $\Delta\pi$ 		& $\Lambda K$	& $\Sigma K$ 	\\ \hline
$I=\oh$         & $\sqrt{3} $	& $1$ 	& $\sqrt{3}$ 	& $-\sqrt{2}$ 		& $1$		& $-\sqrt{3}$	\\
$I=\frac{3}{2}$ & $1$       	& $0$ 	& $1$        	& $\sqrt{\frac{5}{3}}$ 	& $0$ 		& $-1$ 		\\ 
\hline \hline
\end{tabular}
\end{center}
\label{isospin_res}
\end{table}

In \ref{sec:barev} we list the partial wave projected, bare resonance annihilation vertices $v$. The bare
coupling constants obtained in the fit can be found in Table~\ref{bare_cou}. Every vertex function is
multiplied with the corresponding form factor of Eq.~(\ref{fofa}) and the isospin factor $I_R$ listed in
Table~\ref{isospin_res}, 
\be
\gamma_B=F(k)\,I_R\,v\,\sqrt{\frac{E + m_B}{E \, \omega}}
\label{bare_all}
\ee
with $\gamma_B$ from Eq.~(\ref{dressed}). The resonance creation vertices $\gamma_B^{(\dagger)}$ are given
by $\gamma_B^{(\dagger)}=(\gamma_B)^{\dagger}$. In Eq.~(\ref{bare_all}) and \ref{sec:barev}, $E$ and
$\omega$ denote the baryon and meson on-shell c.m. energies, respectively, $m_B$ is the baryon mass of the
channel, and $k=|\vec{k}|$ is the baryon-meson c.m. momentum. 

As mentioned in Sec.~\ref{sec:schannel}, the resonance vertices with total spin $J\geq5/2$ have not been
derived from Lagrangians. Instead, they have been constructed obeying the  correct dependence on the orbital
angular momentum $L$ (centrifugal barrier). From parity considerations, one can easily relate the bare
vertices from the resonances with $J\le 3/2$ to those of higher spin,
\begin{align}
(\gamma_B)^{\rm MB}_{\frac{5}{2}^{-}}	 &=\frac{k}{m_B}\;(\gamma_B)^{\rm MB}_{\frac{3}{2}^{+}}	  &	  
(\gamma_B)^{\rm MB}_{\frac{7}{2}^{+}}	 &=\frac{k^{2}}{m_B^{2}}\;(\gamma_B)^{\rm MB}_{\frac{3}{2}^{+}} \non
(\gamma_B)^{\rm MB}_{\frac{5}{2}^{+}}	 &=\frac{k}{m_B}\;(\gamma_B)^{\rm MB}_{\frac{3}{2}^{-}}	 &	 
(\gamma_B)^{\rm MB}_{\frac{7}{2}^{-}}	 &=\frac{k^{2}}{m_B^{2}}\;(\gamma_B)^{\rm MB}_{\frac{3}{2}^{-}} 
\label{higher1}
\end{align}
with the $(\gamma_B)^{\rm MB}_{\frac{3}{2}^{\pm}}$ from Eq.~(\ref{bare_all}). Eq.~(\ref{higher1}) provides
the correct dependence on $L$ for all channels ${\rm MB}=\, \pi N$, $\rho N$, $\pi\Delta$, and $K\Sigma$. 


\subsection{Partial wave projected resonance vertex functions}
\label{sec:barev}
In the following, the classification of vertices corresponds to the quantum numbers of the $\pi N$ channel.
E.g., $S_{11}(S_{31})$ refers to the $I=J=1/2$ resonances that couple to $\pi N$ in $S$-wave, as for example
the $N^*(1535)S_{11}$ or $\Delta(1620)S_{31}$. The other channels can, of course, couple with different
orbital momentum $L$ and $S$ (note there are three $\rho N$ and two $\pi\Delta$ channels with different
combinations of $L$ and $S$). 

In all cases, the vertex functions $v$ for the resonance couplings to $KY$ and $\eta N$ are the same
as for the $\pi N$ case except  for the different masses and isospin coefficients from
Table~\ref{isospin_res}. Thus, the $\eta N,\, KY$ vertices are not quoted explicitly.

\underline{$S_{11}$  $(S_{31})$}
\begin{itemize}
\item $N\pi$ $$
v=-\,i\,\frac{f}{m_\pi}\,\frac{1}{\sqrt{8}\,\pi}\,\left(\omega_\pi + E_N - m_N\right)
$$

\item $N\rho$ (L=0, S=1/2) $$
v=-\,f\,\frac{1}{\sqrt{24}\,\pi}\,\frac{1}{m_\rho}\left(\omega_\rho + E_N - m_N +
2\,m_\rho\right) $$

\item $N\rho$ (L=2, S=3/2) $$
v=-\,f\,\frac{1}{\sqrt{12}\,\pi}\,\frac{1}{m_\rho}\left(\omega_\rho + E_N - m_N -
m_\rho\right) $$

\item $\Delta\pi$ $$
v=-\,i\,\frac{f}{m_\pi}\,\frac{1}{\sqrt{12}\,\pi}\,\frac{E_\Delta + \omega_\pi
}{m_\Delta} \left(E_\Delta - m_\Delta\right) $$

 \end{itemize}

\newpage

\underline{$P_{11}$ $(P_{31})$}
 \begin{itemize} 
\item $N\pi$ $$
v=i\,\frac{f}{m_\pi}\,\frac{1}{\sqrt{8}\,\pi}\,k\,\left(1 + \frac{\omega_\pi}{E_N
+ m_N}\right) $$

\item $N\rho$ (L=1, S=1/2) $$
v=f\,\frac{1}{\sqrt{24}\,\pi}\,\frac{k}{m_\rho}\,\left(1 +
\frac{\omega_\rho}{E_N + m_N} - \frac{2\,m_\rho}{E_N + m_N}\right)
$$

\item $N\rho$ (L=1, S=3/2) $$
v=f\,\frac{1}{\sqrt{12}\,\pi}\,\frac{k}{m_\rho}\,\left(1 +
\frac{\omega_\rho}{E_N + m_N} + \frac{m_\rho}{E_N + m_N}\right) $$

\item $\Delta\pi$ $$
v=i\,\frac{f}{m_\pi}\,\frac{1}{\sqrt{12}\,\pi}\,\frac{k}{m_\Delta}\,\left(E_\Delta +
\omega_\pi\right) $$
\end{itemize}

\underline{$P_{13}$ $(P_{33})$}
\begin{itemize}
\item $N\pi$ $$
v=-\,i\,\frac{f}{m_\pi}\,\frac{1}{\sqrt{24}\,\pi}\,k $$

\item $N\rho$ (L=1, S=1/2) $$
v=\frac{f}{m_\rho}\,\frac{1}{\sqrt{72}\,\pi}\,\frac{k}{E_N+m_N}\left(E_N + m_N -
\omega_\rho - m_\rho\right) $$

\item $N\rho$ (L=1, S=3/2) $$
v=\frac{f}{m_\rho}\,\frac{1}{\sqrt{360}\,\pi}\,\frac{k}{E_N+m_N}\left(5\,E_N +
5\,m_N + 4\,\omega_\rho + m_\rho\right) $$

\item $N\rho$ (L=3, S=3/2) $$
v=\frac{f}{m_\rho}\,\frac{1}{\sqrt{40}\,\pi}\,\frac{k}{E_N+m_N}\left(\omega_\rho
- m_\rho\right) $$

\item $\Delta\pi$ (L=1) $$
v=i\,\frac{f}{m_\pi}\,\frac{1}{\sqrt{360}\,\pi}\,\frac{k}{m_\Delta}\,\left( E_\Delta +
4\,m_\Delta \right)\,\left( 1+ \frac{\omega_\pi}{E_\Delta +
m_\Delta} \right) $$

\item $\Delta\pi$ (L=3) $$
v=-\,i\,\frac{f}{m_\pi}\,\frac{1}{\sqrt{40}\,\pi}\,\frac{k}{m_\Delta}\,\left( E_\Delta - m_\Delta
\right)\,\left( 1+ \frac{\omega_\pi}{E_\Delta + m_\Delta} \right) $$

\end{itemize}

\newpage

\underline{$D_{13}$  $(D_{33})$}
 \begin{itemize}
\item $N\pi$ $$
v=\frac{f}{m_\pi^2}\,\frac{1}{\sqrt{24}\,\pi} \left(E_N-m_N \right) \left(\omega_\pi+E_N+m_N \right)$$

\item $N\rho$ (L=0, S=3/2) $$
v=-\,i\,\frac{f}{m_\rho}\,\frac{1}{\sqrt{72}\,\pi}\,\left(2\omega_\rho + m_\rho + E_N - m_N\right) $$

\item $N\rho$ (L=2, S=1/2) $$
v=-\,i\,\frac{f}{m_\rho}\,\frac{1}{\sqrt{72}\,\pi}\,\left(\omega_\rho - m_\rho - E_N + m_N\right) $$

\item $N\rho$ (L=2, S=3/2) $$
v=-\,i\,\frac{f}{m_\rho}\,\frac{1}{\sqrt{72}\,\pi}\,\left(\omega_\rho - m_\rho + 2 E_N - 2 m_N\right) $$

\item $\Delta\pi$ (L=0) $$
v=\frac{f}{m_\pi}\,\frac{1}{\sqrt{72}\,\pi}\,\frac{1}{m_\Delta}\,\left(\omega_\pi + E_\Delta
-m_\Delta\right) \left( E_\Delta + 2\,m_\Delta \right) $$

\item $\Delta\pi$ (L=2) $$
v=-\,\frac{f}{m_\pi}\,\frac{1}{\sqrt{72}\,\pi}\,\frac{1}{m_\Delta}\,\left(\omega_\pi + E_\Delta
-m_\Delta\right)\left(E_\Delta - m_\Delta \right)
$$

\end{itemize}


\section{Residues and branching ratios}
\label{sec:residues}
The residue $a_{-1}$ and constant term $a_0$ from the Laurent expansion of Eq.~(\ref{pa}) can be obtained by
a closed contour integration along a path $\Gamma (z)$ around the pole position $z_0$,
\be
a_n&=&\frac{1}{2\pi i}\oint_{\Gamma (z)} \frac{T^{(2)}(z)\,dz}{(z-z_0)^{n+1}}.
\label{contourint}
\ee
Alternatively, the residue and subsequent terms in the Laurent expansion can be obtained by an iterative
procedure according to
\be
\frac{\partial}{\partial z}|_{z=z_0} \,\frac{1}{T^{(2)}(z)}		&=&\frac{1}{a_{-1}}\non
\frac{\partial^2}{\partial z^2}|_{z=z_0} \,\frac{1}{T^{(2)}(z)}	&=&-\frac{2a_0}{a_{-1}^2}\non
\frac{\partial^3}{\partial z^3}|_{z=z_0} \,\frac{1}{T^{(2)}(z)}	&=&\frac{6(a_0^2-a_{-1}a_1)}{a_{-1}^3}\non
\ee
which is numerically stable (the inverse $T^{(2)}$ matrix has a simple zero at $z=z_0$) and fast (no
integration required).

The residue and constant term, $a_{-1}$ and $a_{0}$, can be expressed in terms of dressed
quantities~\cite{Doring:2009bi},
\be
a_{-1}&=&\frac{\Gamma_D\,\Gamma_D^{(\dagger)}}{1-\frac{\partial}{\partial
    z}\Sigma}\non
a_0&=&\frac{a_{-1}}{\Gamma_D\,\Gamma_D^{(\dagger)}}\,\left(
  \frac{\partial}{\partial
    z}(\Gamma_D\,\Gamma_D^{(\dagger)})+\frac{a_{-1}}{2}\,\frac{\partial^2}{\partial
    z^2}\,\Sigma\right)  
    +T^{\npo, (2)}(z_0).
\label{resasga}
\ee
where $\Gamma_D$ ($\Gamma_D^{(\dagger)}$, $\Sigma$) is the dressed annihilation vertex (creation vertex,
self-energy) as defined in Eq.~(\ref{dressed}), evaluated on the second
sheet at $z_0$.  Eq. (\ref{resasga}) shows that there is a contribution to the constant term $a_0$ from
$T^\npo$, as expected, but also from the pole term. This is one of the reasons why an identification of
$T^\npo$ as background is problematic~\cite{Doring:2009bi}.

For the two-resonance case shown in Eq.~(\ref{2res}), the residues can be expressed in terms of dressed
quantities like in the one-resonance case of Eq. (\ref{resasga}), 
\be
a_{-1, i}=\left[\frac{\det D^{(2)}}{d/dz\,\det D^{(2)}}\,T^{\po,(2)}\right]_{z\to z_{0,i}}
\label{resi2res}
\ee
where $a_{-1,i}$ is the residue of resonance $i=1,2$ with pole at $z=z_{0,i}$. 

Using Eqs. (\ref{taut}) and (\ref{pa}), the pole residues $r=|r|e^{i\theta}$ as quoted by the PDG~\cite{pdg}
can be calculated.  For the residue phase $\theta$~\cite{pdg} we consider the usual \cite{Hoehler1,Hoehler2}
definition given by
\be
\tau=\tau_B+\frac{|r|\,e^{i\theta}}{M-z-i\Gamma/2}
\label{usualphi}
\ee
for a resonance with width $\Gamma$ on top of a background $\tau_B$. Comparing Eq. (\ref{usualphi}) with Eq.
(\ref{pa}) and  using Eq. (\ref{taut}), the pole residue $r$ and its phase are given by
\be
|r|= |a_{-1}\,\rho_{\pi N}|,\quad\quad
\theta = -\pi+\arctan\left[\frac{{\rm Im}\,(a_{-1}\,\rho_{\pi N})}{{\rm Re}\,(a_{-1}\,\rho_{\pi N})}\right]
\label{rerere}
\ee
where $\rho_{\pi N}$ is the phase space factor $\rho$ from Eq. (\ref{taut}) for the $\pi N\to\pi N$
transition, evaluated at the complex pole position. For the corresponding quantity in the reaction
$\pi^+p\to K^+\Sigma^+$, one simply replaces $\rho_{\pi N}\to \sqrt{\rho_{\pi N}\,\rho_{K\Sigma}}$.

It is convenient to express the $n^2$ different residues $a_{-1}^{i\to f}$ [with $i,f=1,\cdots,n$ for the
transitions within $n$ channels] in terms of $n$ parameters $g$. Indeed, the residues factorize with respect
to the channel space and, e.g., for the residues into the $\pi N$ and $K\Sigma$ channels,
\be
a_{-1}^{i\to f}=g_i\,g_f
\label{couplings}
\ee
with a unique set of $g_i$ up to one undetermined global sign. 
For the channels $\pi N$ and $K\Sigma$, the partial decay widths are evaluated using
\be
\Gamma_f(M_R,M_f,m_f)= \frac{|\tilde{g}_f|^2}{2\pi}\frac{M_f}{M_R}\,k,\quad
\tilde{g}_f=	2\pi\,g_f\,\sqrt{\frac{\omega_f E_f}{M_f}}
\label{brastable}
\ee
with the resonance mass (final baryon, final meson mass) $M_R={\rm Re}\,z_0$ ($M_f,\,m_f$) and $\omega_f$,
$E_f$ are the meson and nucleon energy at the on-shell momentum $k$. 

The sum of partial decay widths should equal the total width, $\sum_f\Gamma_f=\Gamma_{\rm tot}$. The
right-hand side of this equation can be determined independently ($\Gamma_{\rm tot}=-2\,i\,{\rm Im}\,z_0$)
and be used as a test of the formalism. Indeed, below the $\pi\pi N$ threshold, the equality holds to the 1
\% level (The definition of branching ratios into the effective $\pi\pi N$ states $\rho N$, $\sigma N$,
$\pi\Delta$ is model-dependent anyways). Although Eq.~(\ref{brastable}) is a good approximation to the
partial decay widths, it should be noted that $\sum_f\Gamma_f=\Gamma_{\rm tot}$ never holds exactly, even in
a manifestly unitary coupled-channels model with only stable intermediate states. This is simply because the
amplitude has non-analytic branch points, required by unitarity, and this information is not contained in
the $g$. However, this does not become a real issue unless a pole is very close to a branch point.


\end{document}